\documentclass[3p,12pt]{elsarticle}
\usepackage{threeparttable,booktabs,tabularx}
\usepackage[fleqn]{amsmath}
\usepackage{cases}
\usepackage{multirow}
\usepackage{xcolor}
\usepackage[normalem]{ulem}
\usepackage{tabularx}
\usepackage{siunitx}
\usepackage[caption=false]{subfig}
\usepackage{comment}
\usepackage{algorithm}
\usepackage{algpseudocode}
\usepackage{algorithmicx}
\usepackage{grffile}
\usepackage{rotating}
\usepackage{array}
\usepackage{lineno}
\usepackage{hyperref}
\usepackage{makecell}
\usepackage{graphicx,enumerate}
\usepackage{amssymb}
\usepackage{bm,amsmath}
\usepackage[caption=false]{subfig}
\usepackage{caption,color}
\usepackage{threeparttable}
\usepackage{subeqnarray}
\usepackage{multirow}
\usepackage{lscape}
\usepackage{rotating}
\usepackage{graphics}
\floatname{algorithm}{Algorithm}
\usepackage{epstopdf}
\journal{.}

\allowdisplaybreaks

\biboptions{numbers,sort&compress} % 压制引用，[1,2,3]--> [1-3]

\begin{document}
\begin{frontmatter}

\title{Multiscale simulation of rarefied polyatomic gas flow via DIG method}

\author{Liyan Luo \fnref{Equal}}
\author{Tao Huang \fnref{Equal}}
\author{Qi Li\corref{corresponding}}
\ead{liq33@sustech.edu.cn}
\author{Lei Wu}

\fntext[Equal]{These authors contribute equally to this work}
\cortext[corresponding]{Corresponding author}

\address{Department of Mechanics and Aerospace Engineering,
Southern University of Science and Technology, 518055 Shenzhen, China}

\begin{abstract}
A novel multiscale numerical method is developed to accelerate direct simulation Monte Carlo (DSMC) simulations for polyatomic gases with internal energy. This approach applies the general synthetic iterative scheme to stochastic simulations, removing the inherent restrictions on spatial grid size in DSMC and boosting the evolution of the particle distribution towards a steady state. Firstly, the proposed method intermittently couples the standard DSMC solver with the solution of steady-state macroscopic synthetic equations derived from the kinetic equation for polyatomic gases. These synthetic equations, encompassing accurate constitutive relations and relaxation rates, ensure applicability across the entire flow regime. Secondly, the particle distribution within the DSMC framework is adjusted to align with the solution of the macroscopic equations, significantly accelerating convergence to the steady-state solution. Finally, an adaptive treatment is implemented for the constitutive relations. Higher-order terms extracted from DSMC are applied only in rarefied regions where the local Knudsen number exceeds a predefined threshold. The fast convergence and asymptotic-preserving properties of the proposed method are demonstrated through several numerical simulations, in which it achieves remarkable speedups compared to standard DSMC, particularly in the near-continuum regime.
\end{abstract}

\begin{keyword}
rarefied gas dynamics, direct simulation Monte Carlo, general synthetic iterative scheme
\end{keyword}

\end{frontmatter}
\section{Introduction}\label{sec:1}
Rarefied gas dynamics is of paramount importance in diverse engineering fields, including aerospace applications such as hypersonic flight~\cite{ivanov-1998,Anderson2019}, as well as advanced technologies in lithography systems~\cite{Wang2024JVSTB} and microelectromechanical systems~\cite{titov-2008}. The degree of gas rarefaction is quantified by the Knudsen number ($\text{Kn}$), defined as the ratio of the mean free path of gas molecules to a characteristic length. For near-continuum flows ($\text{Kn}<0.1$), the Navier--Stokes--Fourier (NSF) equations provide adequate predictions by taking into account the rarefaction effects through the boundary conditions of velocity slip and temperature jump. However, as $\text{Kn}$ increases, NSF solutions deviate substantially from experimental observations, since the constitutive relations of Newtonian viscosity and Fourier thermal conduction break down. Consequently, the Boltzmann equation, which describes the statistical behavior of gas molecules at the kinetic level, is essential for accurately modeling and predicting the behavior of rarefied gases across the entire spectrum of rarefaction. 

Two types of numerical methods are prevailing in simulating rarefied gas flows: the deterministic discrete velocity method (DVM)~\cite{aristov-2001} and the stochastic direct simulation Monte Carlo (DSMC) method~\cite{bird-1994}. In DVM approaches, the velocity distribution function of gas molecules is discretized in both spatial and velocity domains, enabling direct numerical solutions using sophisticated computational fluid dynamics techniques. In contrast, the DSMC method models gas flows by representing the gas as a collection of simulation particles, typically requiring only a few dozen particles per spatial cell. These particles move freely between collisions, which are modeled stochastically based on collision rates and post-collisional velocities determined by kinetic theory. However, both conventional DVM and DSMC approaches suffer from unaffordable computational cost in the near-continuum flow regime. This limitation arises primarily from two factors: (i) the requirement for grid sizes and time steps finer than the molecular mean free path and mean collision time, respectively, due to the splitting treatment of particle streaming and collisions; and (ii) the inefficient information exchange across length scales significantly larger than the mean free path, resulting from the localized nature of molecular collisions~\cite{Wang2018CF}. Therefore, the inherent multiscale nature of numerous practical problems, arising from complex geometries with varying length scales, and the diverse relaxation processes (e.g., internal energy, chemical reactions), necessitates the development of multiscale methodologies for efficiently and accurately solving the kinetic equation.

Over the past decade, remarkable achievements have been made in the development of multiscale schemes based on deterministic methods. For example, the unified gas kinetic scheme~\cite{Xu2010JCP} and its discrete version~\cite{Guo2013PRE} solve the convection and collision simultaneously, thereby relieving the constraints of the mean free path on the grid size and asymptotically preserving the NSF solutions. The general synthetic iterative scheme (GSIS)~\cite{su-2020,su-20201} alternately solves the macroscopic synthetic equation and mesoscopic kinetic equation, not only demonstrating minimal numerical dissipation with coarse spatial grids but also exhibiting fast convergence across all flow regimes. Nevertheless, despite their success in simulating many multiscale problems~\cite{zhu-2016,zhang-2024,zeng-2023,shi-2024}, the DVM-based schemes encounter difficulties when applied to problems involving strong non-equilibrium phenomena (e.g., hypersonic flows). In these cases, a substantial increase in the number of velocity grid points becomes necessary, leading to massive demands on computational resources, including memory and processing time. Moreover, constructing accurate kinetic models for complex physicochemical processes presents a significant challenge, hindering the extension of these methods within the DVM framework to such intricate problems.

Many strategies have also been proposed to enhance the efficiency of stochastic methods in the near-continuum flow regime. Hybrid NSF--DSMC algorithms~\cite{schwartzentruber-2005,schwartzentruber-2007} partition the simulation domain into continuum and rarefied regions, applying NSF equations and the DSMC method, respectively, but face significant challenges in accurately defining the domain interface. To circumvent this, the time-relaxed Monte Carlo method~\cite{pareschi-2001} and the asymptotic-preserving Monte Carlo method~\cite{ren-2014} have been developed, although they preserve only the Euler limit in the continuum regime~\cite{Jin1999SIAM}. Recently, a time-relaxed Monte Carlo method that accurately preserves the NSF equations has been established~\cite{Fei2023JCP}, where micro-macro decomposition of the Boltzmann collision operator is based on the Chapman--Enskog expansion. Alternately, hybrid approaches that couple macroscopic solvers with stochastic methods offer powerful multiscale strategies. The moment-guided Monte Carlo method~\cite{degond-2010} utilizes solutions of moment equations to guide the evolution of particle distributions in DSMC. The unified gas-kinetic wave-particle methods~\cite{liu-2019} decompose the gas distribution into equilibrium and non-equilibrium components within each spatial cell, solving for the equilibrium part with a macroscopic solver and tracking the non-equilibrium part using the Monte Carlo approach. Additionally, the hybrid GSIS-DSMC method~\cite{Luo2024arXiv} have been developed, building upon the NSF-preserving time-relaxed Monte Carlo approach~\cite{Fei2023JCP} and the GSIS algorithm~\cite{luo-2023}. This method enhances the asymptotic-preserving characteristic of the stochastic solver and substantially accelerates the convergence of DSMC simulations towards steady-state solutions.

However, developing NSF limit asymptotic-preserving and time-relaxed schemes for the DSMC method requires tremendous mathematical skills and complex implementation procedures. Therefore, extending such approaches to problems involving intricate physicochemical relaxation processes remains a considerable challenge and may introduce additional computational costs. To address these limitations, we recently developed the Direct Intermittent GSIS--DSMC (DIG) method~\cite{luo-2024}, a simpler numerical framework possessing both asymptotic-preserving and fast-convergence properties for efficient and accurate simulations of rarefied gas dynamics. In DIG, the solution of macroscopic synthetic equations is intermittently applied to the DSMC simulation, typically at intervals of 50 to 100 time steps, through a simple linear transformation of particle velocities. The duration of these intermittent intervals is carefully chosen to optimize the balance between computational efficiency and accuracy. Due to its minimal modifications to the standard DSMC framework, DIG is well-suited for accelerating simulations of multiscale gas flows with complex relaxation processes.

In this paper, we develop the DIG method to simulate polyatomic gas flows, incorporating rotational energy alongside translational degrees of freedom. The widely adopted Borgnakke--Larsen collision model~\cite{borgnakke-1975} is employed in DSMC to capture the energy exchange between translational and rotational modes. The macroscopic synthetic equations governing internal energy relaxation, as developed in deterministic GSIS solver for polyatomic gas~\cite{zeng-2023}, are integrated into the DIG framework. Additionally, an adaptive treatment of constitutive relations is implemented, applying higher-order terms extracted from DSMC only in locally rarefied regions, further enhancing computational efficiency. It is noted that the inclusion of vibrational modes within this framework is straightforward.

The remainder of this paper is organized as follows. Section~\ref{sec:DSMC_equation} presents the DSMC method for polyatomic gases and the corresponding macroscopic synthesis equations. Section~\ref{sec:DIG} details the DIG algorithm that couples these two solvers. Numerical tests for lid-driven cavity flows and hypersonic flows passing a cylinder of nitrogen gas are conducted in section~\ref{sec:numerical_case}, to assess the accuracy and efficiency of the proposed DIG method. Finally, conclusions and outlooks are summarized in section~\ref{sec:conclusions}.
\section{DSMC and macroscopic equations for polyatomic gas}\label{sec:DSMC_equation}

The DIG method, like GSIS, employs an alternating application of mesoscopic and macroscopic solvers. This section will introduce the DSMC method for polyatomic gases and the corresponding macroscopic synthesis equations that incorporate rarefaction effects.

Note that all variables used in this paper are non-dimensionalized with respect to the following reference quantities: length $L_0$, density $\rho_0$, temperature $T_0$, and speed $\sqrt{RT_0}$, where $R$ is the specific gas constant. Additionally, the global Knudsen number $\text{Kn}$ of a gas flow problem is defined in terms of these reference quantities as,
\begin{equation}
    \text{Kn} = \frac{\mu(T_0)}{p_0 L_0} \sqrt{\frac{\pi R T_0}{2}},
\end{equation}
where $p_0=\rho_0 R T_0$ is the reference pressure, and $\mu(T_0)$ is the shear viscosity of the gas at temperature $T_0$.

\subsection{Kinetics of polyatomic gas}

Wang-Chang and Uhlenbeck~\cite{WangCS1951} extended the Boltzmann equation to describe polyatomic gases by incorporating internal degrees of freedom quantum mechanically. An individual velocity distribution function $f_i(t,\bm{x},\bm{v})$ is assigned to each internal energy level $i$, where $t$ represents the time, $\bm{x}$ is the spatial coordinates and $\bm{v}$ is the translational molecular velocity. In the absence of external forces, the evolution of each velocity distribution function is governed by the following equation,
\begin{equation}\label{eq:WCU_equation}
	\begin{aligned}
		\frac{\partial{f_{i}}}{\partial{t}}+\bm{v} \cdot \frac{\partial{f_{i}}}{\partial{\bm{x}}} = \sum_{i'j'}\sum_{j}\int_{-\infty}^{\infty}\int_{4\pi}{\left(\frac{g_{i}g_{j}}{g_{i'}g_{j'}}f_{i'}f_{j'}-f_{i}f_{j}\right)|\bm{v}-\bm{v}_*|\sigma_{ij}^{i'j'}\mathrm{d}\Omega\mathrm{d}\bm{v}_*},
	\end{aligned}
\end{equation}
where $\bm{v}$ and $\bm{v}_*$ are the pre-collision velocities of the two molecules with energy level $i$ and $j$, respectively, and the superscript $'$ indicates the state after collision; $g_i$ is the degeneracy of energy level $i$, $\sigma_{ij}^{i'j'}$ is the scattering cross-section, and $\Omega$ is the solid angle. 

A unique feature of polyatomic gases is the exchange of energy among different degrees of freedom during binary collisions. A collision is termed elastic when the kinetic energy of the translational mode remains conserved; otherwise, it is inelastic. Inelastic collisions lead to a relaxation process where the rotational temperature $T_r$ approaches equilibrium with the translational temperature $T_t$. This relaxation process can be described by the Jeans--Landau equation,
\begin{equation}\label{eq:Jeans_Landau}
	\frac{\mathrm{d}{T_r}}{\mathrm{d}{t}} = \frac{p}{\mu}\frac{T-T_r}{Z},
\end{equation}
where $Z$ is the rotational collision number, and $T=(3T_t+d_rT_r)/(3+d_r)$ is the total temperature, with $d_r$ representing the rotational degree of freedom. Clearly, the energy relaxation rate, which directly influences the bulk viscosity of a gas, is recovered by the correct selection of the rotational collision number~\cite{wu2022rarefied}.

Meanwhile, the internal modes in polyatomic gas also carry the thermal energy and contribute to the heat transport. Thus, the relaxations of heat fluxes are coupled within all the degree of freedom, which is found to satisfy~\cite{Mason1962JCP},
\begin{equation}\label{eq:heat_flux_relaxation}
	\left[ 
      \begin{array}{cc} 
        \partial{\bm{q}_{t}}/{\partial{t}} \\ \partial{\bm{q}_{r}}/{\partial{t}} 
      \end{array}
    \right]
    = -\frac{p}{\mu}
    \left[ 
      \begin{array}{cc} 
        A_{tt} & A_{tr} \\ A_{rt} & A_{rr} 
      \end{array}
    \right]
    \left[ 
      \begin{array}{cc} 
        \bm{q}_{t} \\ \bm{q}_{r} 
      \end{array}
    \right],
\end{equation}
where $\bm{q}_{t}$ and $\bm{q}_{r}$ are the translational and rotational heat fluxes, respectively, and $\bm{A}$ is a matrix encapsulating the thermal relaxation rates determining the translational and internal thermal conductivities of the polyatomic gases. In general, the values of $\bm{A}$ and $Z$, representing the rates of different relaxation processes, can be independent of each other in real gases.

\subsection{DSMC method with Borgnakke--Larsen model}

In a DSMC simulation, gas molecules are represented by a collection of particles, each representing a large number of actual gas molecules (the number is denoted by $N_{\text{eff}}$). These particles carry information about their velocity $\bm{v}$ and internal energy $I_r$, and transport within the simulation domain discretized into cells. The local macroscopic properties, including density $\rho$, flow velocity $\bm{u}$,  translational and rotational temperatures $T_t$ and $T_r$, shear stress $\bm{\sigma}$, and translational and rotational heat flux $\bm{q}_t$ and $\bm{q}_r$, are then sampled within each computational cell with volume $V_{\text{cell}}$,
\begin{equation}\label{eq:DSMC_macroscopic}
    \begin{aligned}
    &\rho=\frac{N_\text{eff}}{V_\text{cell} }N_{p},\,\, u_i=\frac{1}{N_p}\sum_{p=1}^{N_{p}}v_{i,p},\\
    &\sigma_{ij}= \frac{N_\text{eff}}{V_\text{cell} }\sum_{p=1}^{N_p}\left[\left(v_{i,p}-u_i\right)\left(v_{j,p}-u_j\right)-\frac{\delta_{ij}}{3}\left|\bm{v}_{p}-\bm{u}\right|^2\right],\\
    &T_t=\frac{1}{3N_p}\sum_{p=1}^{N_{p}}\left|\bm{v}_p-\bm{u}\right|^2,\,\, T_r = \frac{1}{d_rN_p}\sum_{p=1}^{N_{p}}I_{r,p}\\
    &q_{t,i}=\frac{N_\text{eff}}{2V_\text{cell} }\sum_{p=1}^{N_p}\left(v_{i,p}-u_i\right)\left|\bm{v}_{p}-\bm{u}\right|^2,\,\,q_{r,i}=\frac{N_\text{eff}}{V_\text{cell} }\sum_{p=1}^{N_p}\left(v_{i,p}-u_i\right)I_{r,p},
    \end{aligned}
\end{equation}
where $N_p$ is the number of simulation particles in the cell, subscript $i,j$ represent the direction, and $\delta_{ij}$ is the Kronecker delta. It is noted that, since the characteristic temperatures of the rotational modes of typical polyatomic molecules are less than 100 K, the rotational degrees of freedom can be considered fully excited at room temperature. This allows for a classical representation where the rotational energy of gas molecules $I_r$ takes on continuous values. 

The core principle of the DSMC method lies in decoupling particle motion and collisions. In each time step, particles move freely according to their velocities. Subsequently, collisions between particles within each computational cell are modeled. The probability of a collision occurring between a pair of particles is determined based on their relative velocity, the local gas density, and the interaction model. Once a collision is selected to have occurred, new post-collisional velocities for the colliding particles must be determined. For elastic collisions, the post-collisional velocities are typically determined by statistically conserving momentum and energy. However, for polyatomic gases, the internal energy of the molecules must also be considered.

Directly employing Eq.~\eqref{eq:WCU_equation} for polyatomic gas simulations presents a significant challenge, since determining the cross-sections for all possible transitions between energy levels is computationally prohibitive. Alternately, phenomenological models provide a computationally efficient way, by incorporating the energy exchange between different modes within a monatomic collision framework. The most widely used phenomenological model is that of Borgnakke--Larsen~\cite{borgnakke-1975}. In this model, when a collision is regarded as inelastic, the total energy, considering both the relative translational energy of the pair and the internal energy of the selected particle, is redistributed between the translational and internal modes based on energy equipartition. The fraction of inelastic collisions $P_{inelastic}$, which is approximately the inverse of the relaxation collision number, effectively determines the energy relaxation rate. However, its exact value depends on the selection procedure of inelastic collision in implementing phenomenological models~\cite{Haas1994POF,Zhang2013POF}. In this work, we employ the particle selection method that permits double relaxation, as proposed by Bird~\cite{bird-1994}. To ensure an accurate representation of the relaxation process, the fraction of inelastic collisions is~\cite{li-2021},
\begin{equation}\label{eq:P_inelastic}
    \begin{aligned}
        P_{inelastic} = \frac{\alpha\left(5-2\omega\right)\left(7-2\omega\right)}{5\left(\alpha+1\right)\left(\alpha+2\right)Z},
    \end{aligned}
\end{equation}
where $\omega$ is the shear viscosity index, and $\alpha$ is the angular scattering parameter in the variable-soft-sphere (VSS) collision model.

However, it should be noted that the phenomenological collision model for polyatomic gases in DSMC primarily focuses on achieving the correct relaxation rate of energy. The thermal conductivity and its underlying relaxation rate of heat flux (the values of $\bm{A}$ in Eq.~\eqref{eq:heat_flux_relaxation}) cannot be simultaneously recovered using this single adjustable parameter $P_{inelastic}$~\cite{Wu2020JFM,li-2021}. 

\subsection{Macroscopic synthetic equations}

To inherit the fast convergence and asymptotic-preserving properties of the GSIS, the macroscopic synthetic equations need to be solved together with the DSMC simulation. By combining conservation equations of mass, momentum, and total energy, along with a relaxation equation of transitional--rotational energy exchange, the governing equations for the conserved macroscopic variables $\rho$, $\bm{u}$, $T_t$ and $T_r$ are obtained~\cite{zeng-2023},
\begin{equation}\label{eq:macroscopic_equation}
    \begin{aligned}
        \frac{\partial \rho}{\partial t} + \nabla \cdot (\rho \bm{u}) &=0,\\
        \frac{\partial (\rho \bm{u})}{\partial t} + \nabla \cdot (\rho \bm{u} \bm{u}) +\nabla \cdot (\rho T_t \bm{I} + \bm{\sigma}) &= 0,\\
        \frac{\partial (\rho e)}{\partial t} + \nabla \cdot (\rho e \bm{u}) +\nabla \cdot (\rho T_t \bm{u} + \bm{\sigma} \cdot \bm{u} + \bm{q}_t + \bm{q}_r) &=0,\\
        \frac{\partial (\rho e_r)}{\partial t} + \nabla \cdot (\rho e_r \bm{u}) + \nabla \cdot \bm{q_r} &= \frac{d_r \rho}{2} \frac{T-T_r}{Z \tau},
    \end{aligned}
\end{equation}
where $e = (3T_t + d_r T_r)/2 + u^2/2$ and $ e_r = d_rT_r/2$ are the specific total and rotational energies, respectively; $\tau=\mu/p$ is the relaxation time.

In rarefied gas flows, the stress $\bm{\sigma}$ and heat fluxes $\bm{q}_t,~\bm{q}_r$ in Eq.~\eqref{eq:macroscopic_equation} cannot be solely expressed as functions of the macroscopic variables $\rho$, $\bm{u}$, $T_t$ and $T_r$. Instead, they comprise two parts~\cite{su-2020},
\begin{equation}\label{eq:full_constitutive}
    \begin{aligned}
        &\bm{\sigma} = \bm{\sigma}^{\text{NSF}} + \text{HoT}_{\bm{\sigma}}, \\
        &\bm{q}_t = \bm{q}_t^{\text{NSF}} + \text{HoT}_{\bm{q}_t}, \\
        &\bm{q}_r = \bm{q}_r^{\text{NSF}} + \text{HoT}_{\bm{q}_r}. \\
    \end{aligned}
\end{equation}
Here, terms with superscript ${\text{NSF}}$ represent the linear constitutive relations obtained from the Chapman--Enskog expansion~\cite{chapman-1970} of the kinetic equation to the first order of Knudsen number (the NSF approximation),
\begin{equation}\label{eq:ns_constitutive}
    \begin{aligned}
        &\bm{\sigma}_{\text{NSF}} = -\mu \left( \nabla \bm{u} + (\nabla \bm{u})^T - \frac{2}{3} (\nabla \cdot \bm{u}) \bm{I} \right), \\
        &\bm{q}_{t,\text{NSF}} = -\kappa_t \nabla T_t, \\
        &\bm{q}_{r,\text{NSF}} = -\kappa_r \nabla T_r,
    \end{aligned}
\end{equation}
where $\kappa_t$ and $\kappa_r$ denote the translational and rotational thermal conductivity, respectively, which are determined by the thermal relaxation rates $\bm{A}$ as~\cite{li-2023}
\begin{equation}\label{eq:kappa_A}
	\left[ 
      \begin{array}{cc} 
        \kappa_t \\ \kappa_r 
      \end{array}
    \right]
	= \frac{\mu}{2}
	\left[ 
      \begin{array}{cc} 
        A_{tt} & A_{tr} \\ A_{rt} & A_{rr} 
      \end{array}
    \right]^{-1}
    \left[ 
      \begin{array}{cc} 
        5 \\ d_r 
      \end{array}
    \right].
\end{equation}
Meanwhile, terms with $\text{HoT}$ in Eq.~\eqref{eq:full_constitutive} are the higher-order terms that account for rarefaction effects. These terms are directly calculated from the particle velocity and internal energy distributions obtained from DSMC,
\begin{equation}\label{eq:hot_constitutive}
    \begin{aligned}
        &\text{HoT}_{\sigma} = \bm{\sigma}^{\text{DSMC}} - \bm{\sigma}^{\text{NSF}}, \\
        &\text{HoT}_{\bm{q}_t} = \bm{q}_t^{\text{DSMC}} - \bm{q}_t^{\text{NSF}}, \\
        &\text{HoT}_{\bm{q}_r} = \bm{q}_r^{\text{DSMC}} - \bm{q}_r^{\text{NSF}}, 
    \end{aligned}
\end{equation}
where $\bm{\sigma}^{\text{DSMC}},~\bm{q}_t^{\text{DSMC}},~\bm{q}_r^{\text{DSMC}}$ are statistically sampled in the DSMC simulations based on Eq.~\eqref{eq:DSMC_macroscopic}, and $\bm{\sigma}^{\text{NSF}},~\bm{q}_t^{\text{NSF}},~\bm{q}_r^{\text{NSF}}$ are calculated according to Eq.~\eqref{eq:ns_constitutive} with the macroscopic properties $\rho$, $\bm{u}$, $T_t$ and $T_r$ sampled from DSMC as well.

It should be emphasized that, when solving the synthetic equations (Eq.~\eqref{eq:macroscopic_equation}) with the full constitutive relations (Eq.~\eqref{eq:full_constitutive}), the higher-order terms are determined from the preceding DSMC steps. This iterative approach ensures that the macroscopic variables are continuously updated until a steady state is reached. As demonstrated in numerical simulations~\cite{luo-2024}, this treatment facilitates fast convergence and maintains asymptotic-preserving properties across the entire range of gas rarefaction.

\section{Algorithm of DIG for polyatomic gas}\label{sec:DIG}

The common feature shared by both the DIG and GSIS methods is the coupling of mesoscopic and macroscopic solvers. In DIG, the DSMC results provide higher-order constitutive relations to close the synthetic equation, while the macroscopic solution guides the evolution of the particle distribution in the DSMC solver. On the other hand, unlike GSIS, the solution of the macroscopic equations in DIG is not interspersed with each mesoscopic calculation but is applied intermittently, in every $N_s$ DSMC step.

The optimal value of $N_s$ requires careful consideration. An excessively large $N_s$ diminishes the benefits of GSIS (fast convergence and asymptotic-preserving), since the DSMC solver, which does not possess these properties, dominates the simulation. Conversely, an excessively small $N_s$ leads to insufficient sampling in the DSMC simulations, resulting in increased statistical noise in the macroscopic quantities used to solve the synthetic equations.

In our previous work~\cite{luo-2024}, we empirically determined $N_s$ to be 50, when the effective time step of the DSMC simulation is roughly equal to the mean collision time. The DIG algorithm for monatomic rarefied gas has been successfully applied to enhance the convergence of DSMC in the near-continuum flow regime, and its asymptotic-preserving property has also been demonstrated~\cite{luo-2024}. For instance, in a hypersonic argon flow passing a cylinder, when $\text{Kn}=0.01$ and Mach number $\text{Ma}=5$, DIG achieved a smooth steady-state solution using only 40,000 cells and 17.3 core$\times$hours of computational time, compared to over 2 million cells and 600 core$\cdot$hours required by a conventional DSMC simulation.

\begin{figure}[t]
	\centering
	\includegraphics[scale=0.6,clip=true]{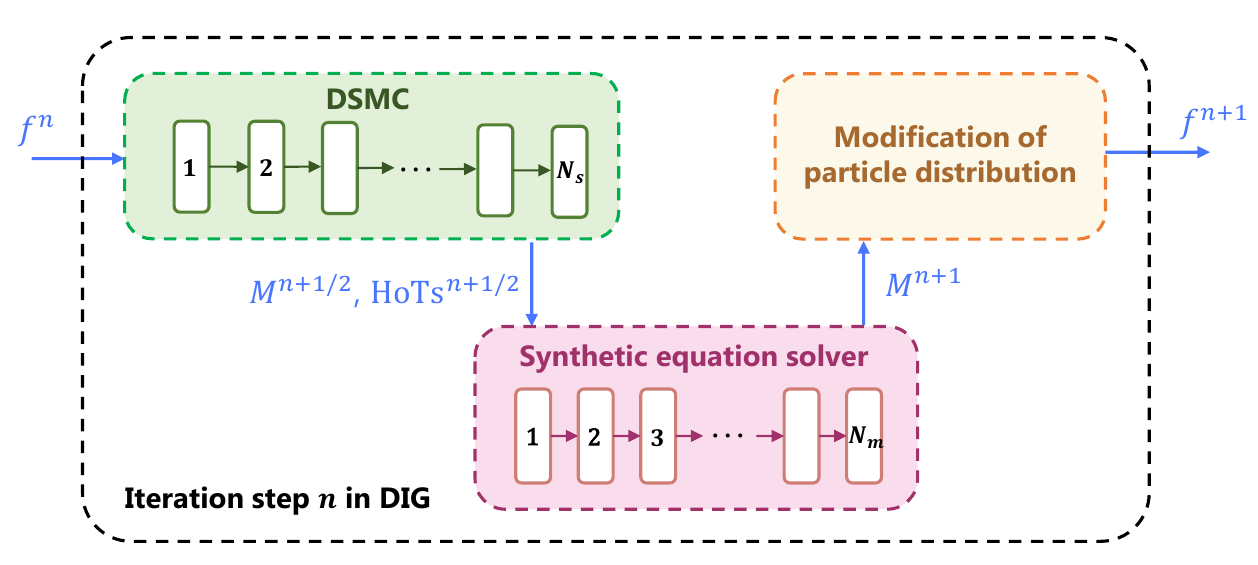}
	\caption{Flowchart of DIG algorithm for polyatomic gases with rotational energy. Each iteration step comprises $N_s$ standard DSMC time steps followed by a solution of the steady-state synthetic equations (with a maximum of $N_m$ inner iterations).} 
	\label{fig:DIG_flowchart}
\end{figure}

For polyatomic gases with rotational energy, the algorithm of DIG, illustrated in Fig.~\ref{fig:DIG_flowchart} comprises the following steps,
\begin{enumerate}
	\item[1.] Initialization: Solve the macroscopic equation (Eq.~\eqref{eq:macroscopic_equation}) with the NSF constitutive relations (Eq.~\eqref{eq:ns_constitutive}). Initialize the simulation particles based on the equilibrium distribution corresponding to the obtained macroscopic properties.
	\item[2.] DSMC solver: In the iteration step $n$, execute the standard DSMC method for $N_s=100$ time steps. Calculate the time-averaged macroscopic quantities $M^{n+1/2}=[\rho^{n+1/2},$ $\bm{u}^{n+1/2}, T_t^{n+1/2}, T_r^{n+1/2}]$ using Eq.~\eqref{eq:DSMC_macroscopic}, and extract higher-order terms $\text{HoTs}^{n+1/2}$ based on Eq.~\eqref{eq:ns_constitutive}.
	\item[3.] Macroscopic equation solver: Given $M^{n+1/2}$ and $\text{HoTs}^{n+1/2}$, solve the synthetic equations (Eq.~\eqref{eq:macroscopic_equation}) with the full constitutive relations (Eq.~\eqref{eq:full_constitutive}) for $N_m=500-2000$ inner iterations, or until the relative error in macroscopic variables between successive iterations falls below $10^{-5}$. Obtain the updated macroscopic quantities $M^{n+1}$. An adaptive treatment of constitutive relations is implemented, and the details are given in section~\ref{sec:adaptive}. The boundary condition and the numerical method for solving the synthetic equation are detailed in Ref.~\cite{zeng-2023,liu-2024}.
	\item[4.] Particle distribution modification: Modify the particle number, velocity, and rotational energy within each computational cell to align with the solution of the synthetic equations $M^{n+1}$. The specific details of this modification step are provided in section~\ref{sec:modifiy_particle}.
\end{enumerate} 
Steps 2-4 are repeated iteratively until the overall solution converges and becomes smooth enough.

\subsection{Adaptive treatment of constitutive relations}\label{sec:adaptive}

The extraction of higher-order terms from DSMC simulations introduces inherent statistical noise due to the stochastic nature of particle-based methods. This statistical fluctuation in the DSMC-derived macroscopic quantities and higher-order terms can significantly impact the stability and accuracy of the macroscopic solver. Given that the higher-order terms are proportional to $\text{Kn}^2$ in the continuum regime and are hence negligibly small~\cite{su-20201}, we introduce an adaptive treatment that the higher-order terms extracted from DSMC are applied only in rarefied regions when the local Knudsen number exceeds a reference value $\text{Kn}_{\text{ref}}$.

To enable adaptive identification of different local flow regimes, we utilize a gradient-length local Knudsen number $\text{Kn}_{\text{GLL}}$ determined by the maximum rate of change of macroscopic variables~\cite{Boyd1995POF},
\begin{equation}
    \text{Kn}_{\text{GLL}} = \lambda\times \max \left( \frac{|\nabla {\rho}|}{\rho}, \frac{|\nabla {u}|}{u}, \frac{|\nabla {T_t}|}{T_t}, \frac{|\nabla {T_r}|}{T_r} \right),
\end{equation}
where $\lambda$ is the local mean free path of gas molecules, and $u=|\bm{u}|$ is the flow speed.

The value of $\text{Kn}_{\text{GLL}}$ in each computational cell measures the local degree of non-equilibrium. Then, the higher-order terms $\text{HoT}_{\sigma}$, $\text{HoT}_{\bm{q}_t}$, $\text{HoT}_{\bm{q}_r}$ in the constitutive relations (Eq.~\eqref{eq:full_constitutive}) are set to zero where $\text{Kn}_{\text{GLL}}<\text{Kn}_{\text{ref}}$. The optimal choice of $\text{Kn}_{\text{ref}}$, requiring a balance of accuracy and efficiency, is generally case-dependent. In this paper, we demonstrate that $\text{Kn}_{\text{ref}} = 0.01$ produces satisfactory outcomes in various numerical simulations. 

\subsection{Modification of particle distribution}\label{sec:modifiy_particle}

In deterministic GSIS schemes, the feedback of updated macroscopic quantities $M^{n+1}$ to the kinetic solver is achieved by explicitly correcting the velocity distribution functions, typically by adjusting their equilibrium parts. However, in DSMC, an explicit expression for the particle distribution function is not available. Therefore, the feedback mechanism in DIG involves modifying the particle number, velocity, and rotational energy to achieve the updated macroscopic quantities $M^{n+1}$ obtained from the macroscopic solver.

The number of simulation particles within each computational cell is first adjusted by applying a replicating and discarding procedure~\cite{degond-2010}, and then the velocity and internal energy of all particles are updated by a simple linear transformation. These procedures subsequently involve the following steps,
\begin{enumerate}
	\item[i.] Determine the target number of particles, $N_p^{n+1}$, based on the updated density $\rho^{n+1}$,
    \begin{equation}\label{}
        \begin{aligned}
            N_p^{n+1}=\text{Iround}\left(\frac{\rho^{n+1} V_{\text{cell}}}{N_{\text{eff}}}\right),
        \end{aligned}
    \end{equation}
    with
    \begin{equation}\label{}
        \begin{aligned}
            \text{Iround}(x)=\begin{cases} \left\lfloor x \right\rfloor+1, & \text{with probability } x-\left\lfloor x \right\rfloor, \\ \left\lfloor x \right\rfloor, &  \text{with probability } 1-x+\left\lfloor x \right\rfloor ,\end{cases},
        \end{aligned}
    \end{equation}
    where $\left\lfloor x \right\rfloor$ represents the largest integer less than or equal to $x$.
	\item[ii.] If $N_p^{n+1}>N_p^{n}$, $N_p^{n+1}-N_p^{n}$ new particles are created. The velocities of these new particles are assigned by randomly selecting existing particles within the cell, and their positions are distributed uniformly over the cell. If $N_p^{n+1}<N_p^{n}$, $N_p^{n}-N_p^{n-1}$ existing particles are randomly selected and removed from the cell.
	\item[iii.] Calculate the temporary velocity $\bm{u}^{*}$ and temperatures $T_t^{*},T_r^{*}$  by sampling the current particle distribution using Eq.~\eqref{eq:DSMC_macroscopic}.
	\item[iv.] Update the velocity and internal energy of each particle using the following linear transformations,
	\begin{equation}\label{}
        \begin{aligned}
            v_i^{n+1} = a v_i^{n} + b_i, \quad I_r^{n+1} = c I_r^{n},
        \end{aligned}
    \end{equation}
    with parameters $a,b_i,c$ determined by,
    \begin{equation}\label{}
        \begin{aligned}
            a = \sqrt{\frac{T_t^{n+1}}{T_t^{*}}}, \quad
            b_i = u_i^{n+1} - u_i^{*} \sqrt{\frac{T_t^{n+1}}{T_t^{*}}}, \quad
            c = \frac{T_r^{n+1}}{T_r^{*}}.
        \end{aligned}
    \end{equation}
\end{enumerate}

\section{Numerical results}\label{sec:numerical_case}

To assess the performance of DIG for polyatomic gases, numerical simulations of two-dimensional nitrogen gas flows are conducted for a lid-driven cavity and a hypersonic flow past a cylinder. The gas--surface interaction is assumed to be fully diffuse in all cases. The collisions of nitrogen molecules in DSMC are described by the VSS model, with the following parameters: mass $m=4.65\times10^{-26}~\text{kg}$, molecular diameter $d=4.11\times10^{-10}~\text{m}$, viscosity index $\omega=0.74$, and angular scattering parameter $\alpha=1$. The rotational degree of freedom is fully excited as $d_r=2$, and the rotational collision number $Z=2.59$ is assumed to be temperature-independent, which corresponds to a fraction of inelastic collisions $P_{inelastic}=0.25$ in DSMC according to Eq.~\eqref{eq:P_inelastic}. The time step applied in the simulations is determined such that it does not exceed the mean collision time. The macroscopic synthetic equations are solved using the finite volume method, with the convention fluxes evaluated implicitly using the lower-upper symmetric Gauss-Seidel technique (further details on the numerical scheme implementation are available in Ref.~\cite{zeng-2023,liu-2024}). The solution of the macroscopic equations is applied every 100 DSMC time steps. All numerical simulations are performed on a parallel computer with the AMD EPYC 7763 processor (2.45GHz) .

\subsection{Lid-driven cavity flow}

Consider a unit square cavity (with side length normalized by $L_0$) with isothermal walls maintained at a temperature of $T_w=1$ (normalized by $T_0$). Simulations are performed for a set of Knudsen numbers ranging from $10^{-4}$ to $0.1$. For $\text{Kn}>0.01$, the top lid of the cavity moves horizontally with a velocity of $U_w=\sqrt{2}$ (normalized by the reference speed $\sqrt{RT_0}$). For $\text{Kn}<0.01$, lower lid velocities ($U_w=0.12$ and $0.25$) are used to avoid turbulence. These velocities correspond to Reynolds numbers $\text{Re}=100$ and $1000$, and $\text{Kn} = 2.63 \times 10^{-3} $ and $ 5.26 \times 10^{-4} $, respectively. The entire simulation domain is discretized into orthogonal grids. Each computational cell is initially populated with an average of 100 simulation particles. The particle velocities are sampled from an equilibrium distribution function, assuming unit density, temperature, and zero flow velocity.

\begin{figure}[h]
    \centering
    \begin{tabularx}{\textwidth}{*{3}{X}} 
        \subfloat{
            \includegraphics[width=0.95\linewidth,trim={30 20 30 60},clip]{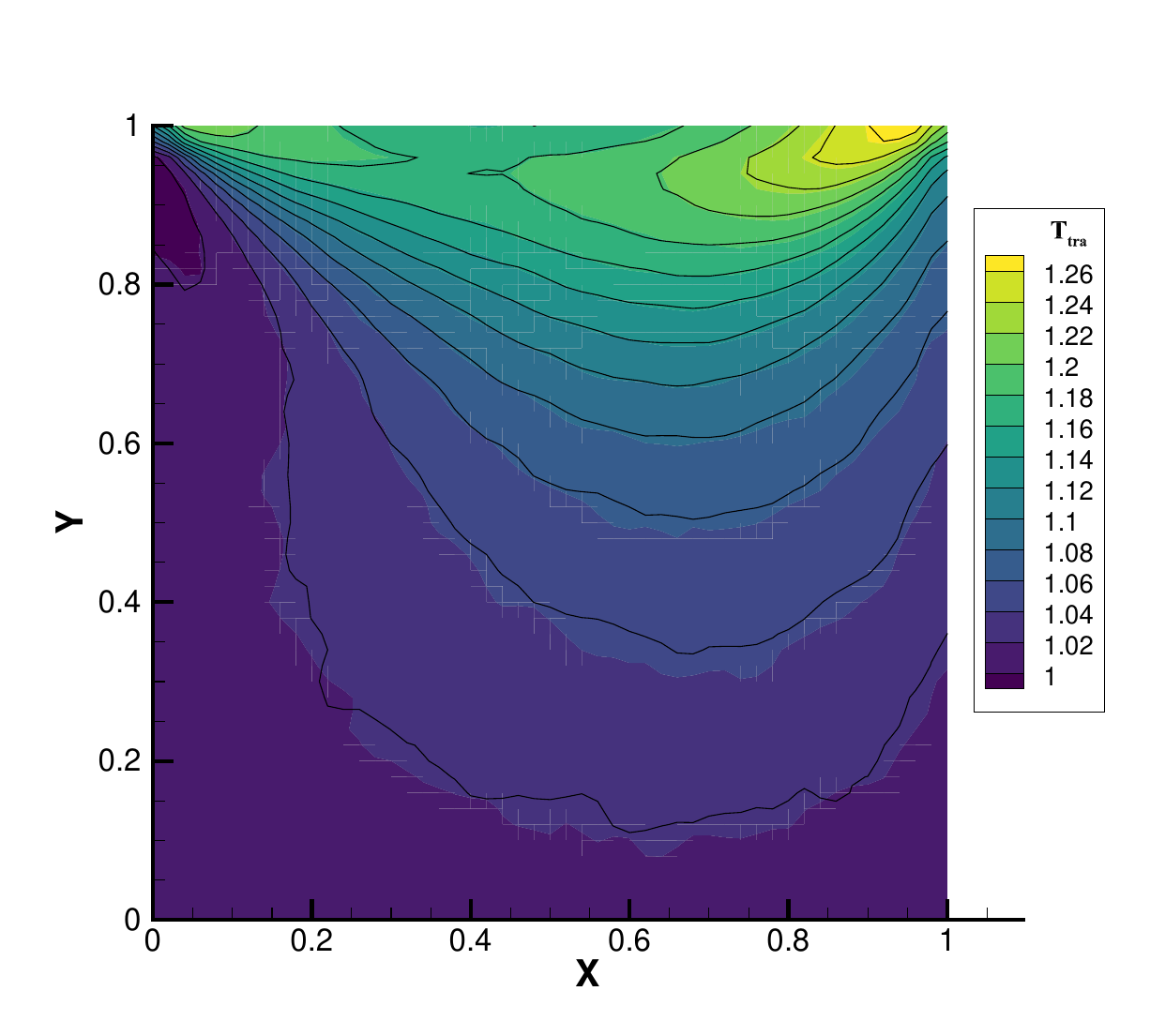}
        } &
        \subfloat{
            \includegraphics[width=0.95\linewidth,trim={30 20 30 60},clip]{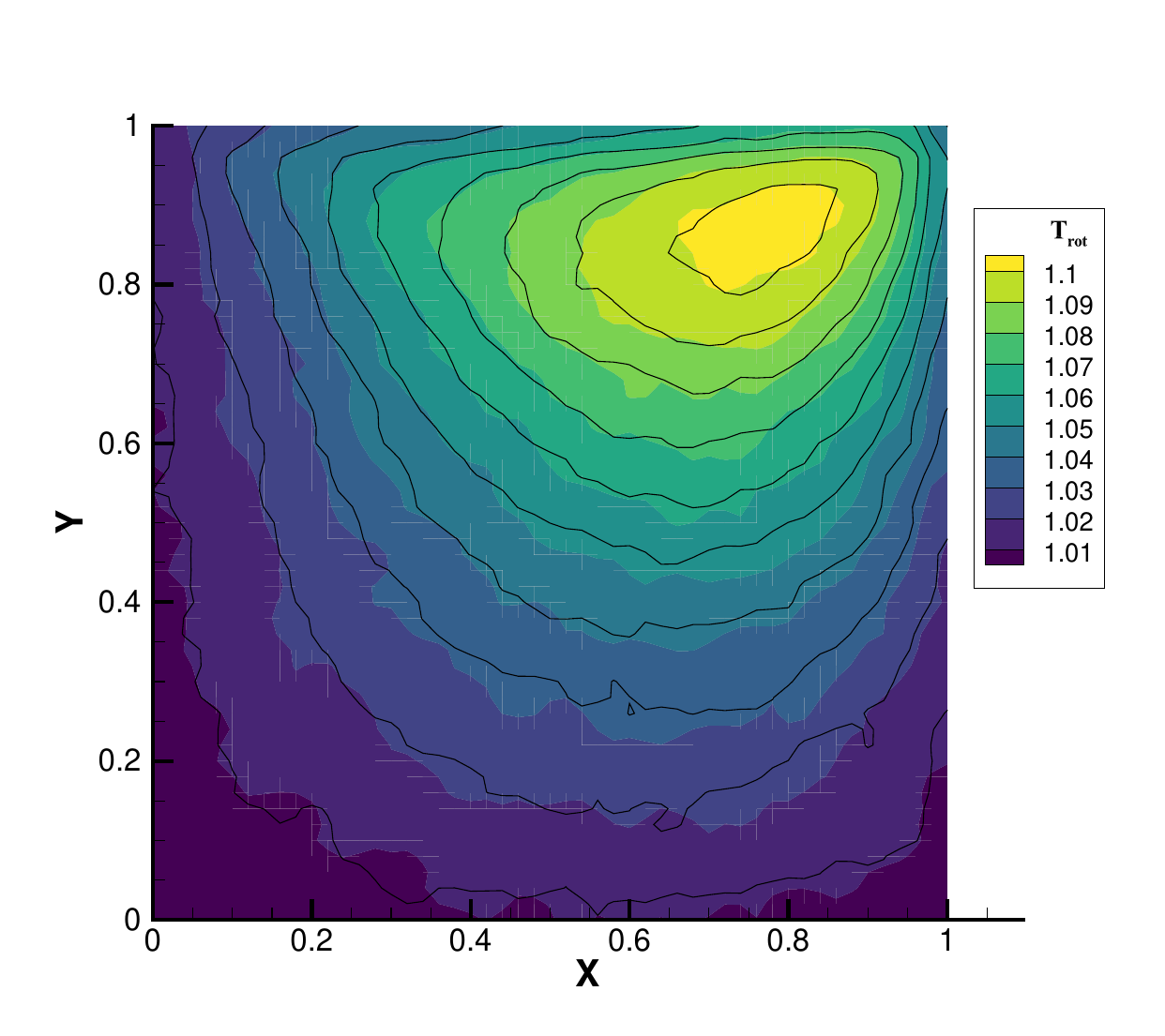}
        } & 
        \subfloat{
            \includegraphics[width=0.95\linewidth,trim={30 20 30 60},clip]{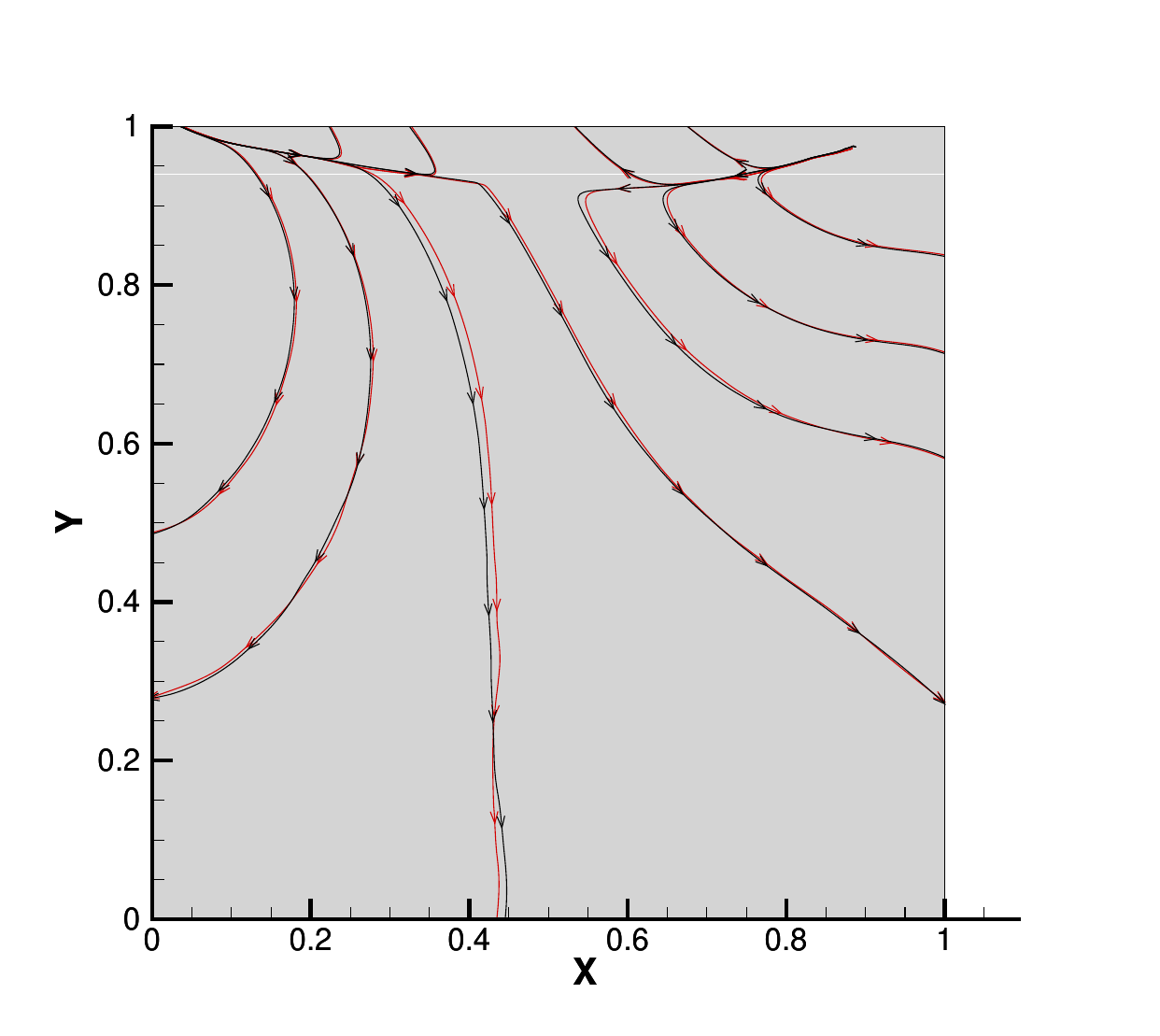}
        } 
        \\
        \subfloat{
            \includegraphics[width=0.95\linewidth,trim={30 20 30 60},clip]{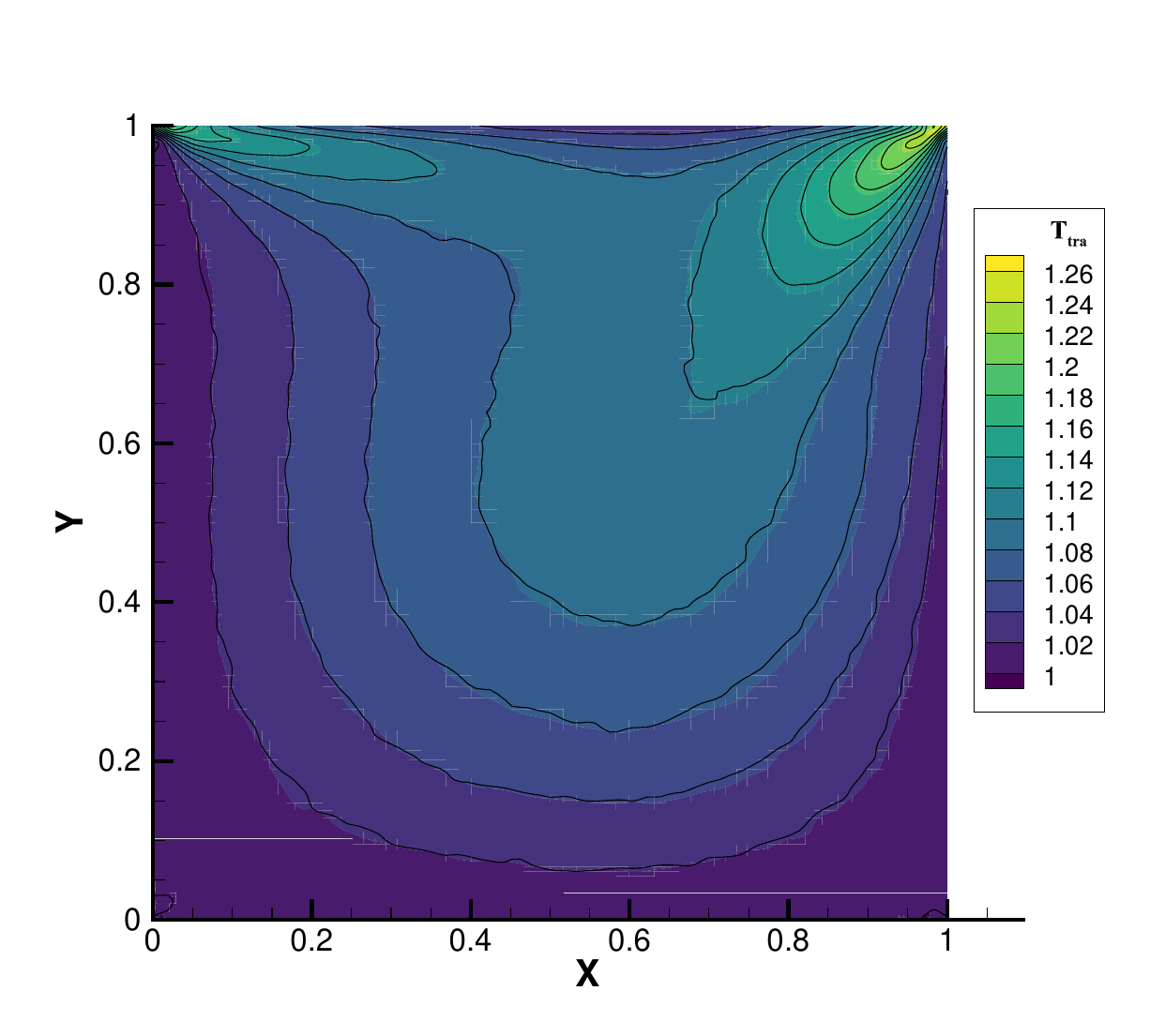}
        } &
        \subfloat{
            \includegraphics[width=0.95\linewidth,trim={30 20 30 60},clip]{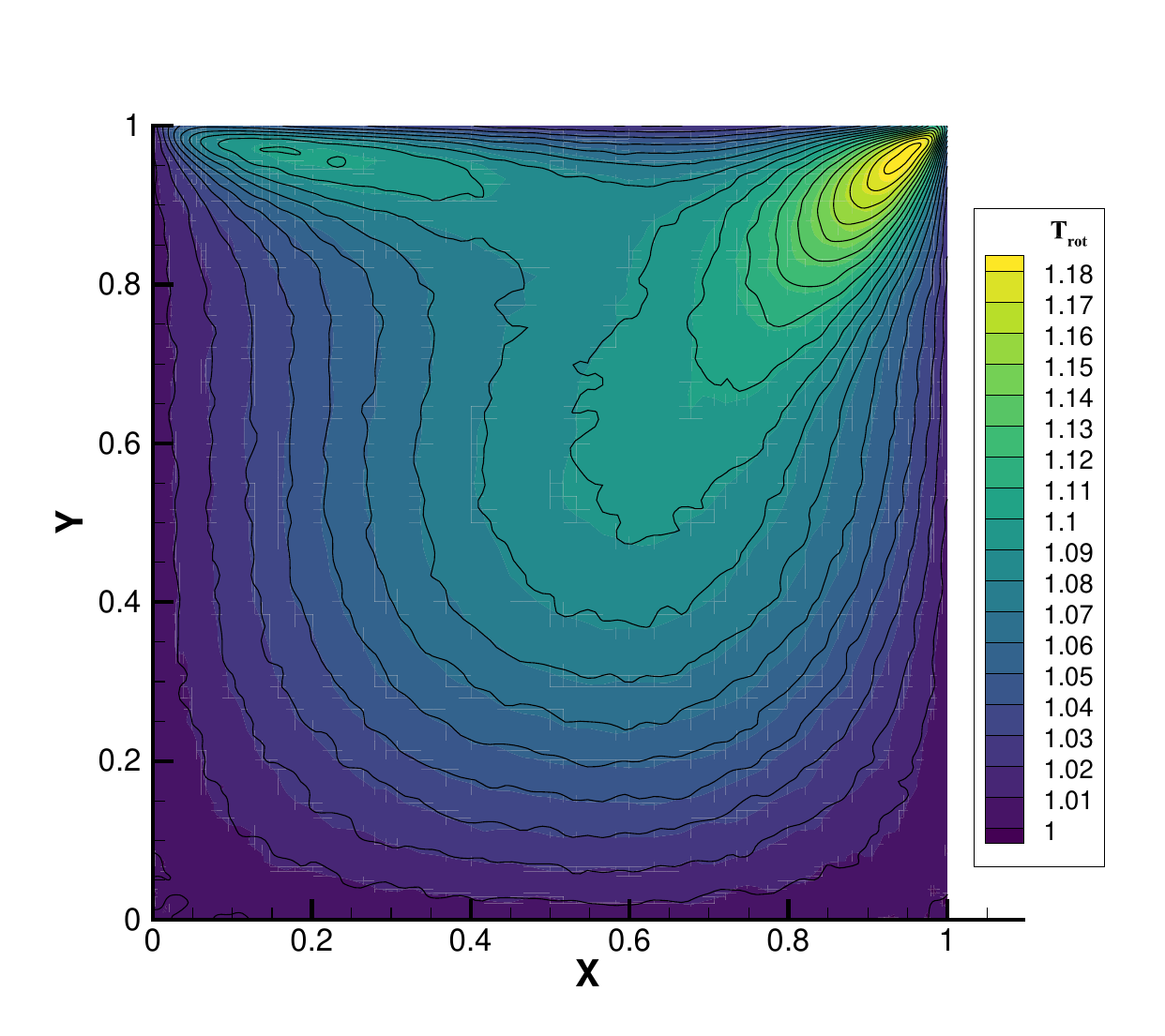}
        } &
        \subfloat{
            \includegraphics[width=0.95\linewidth,trim={30 20 30 60},clip]{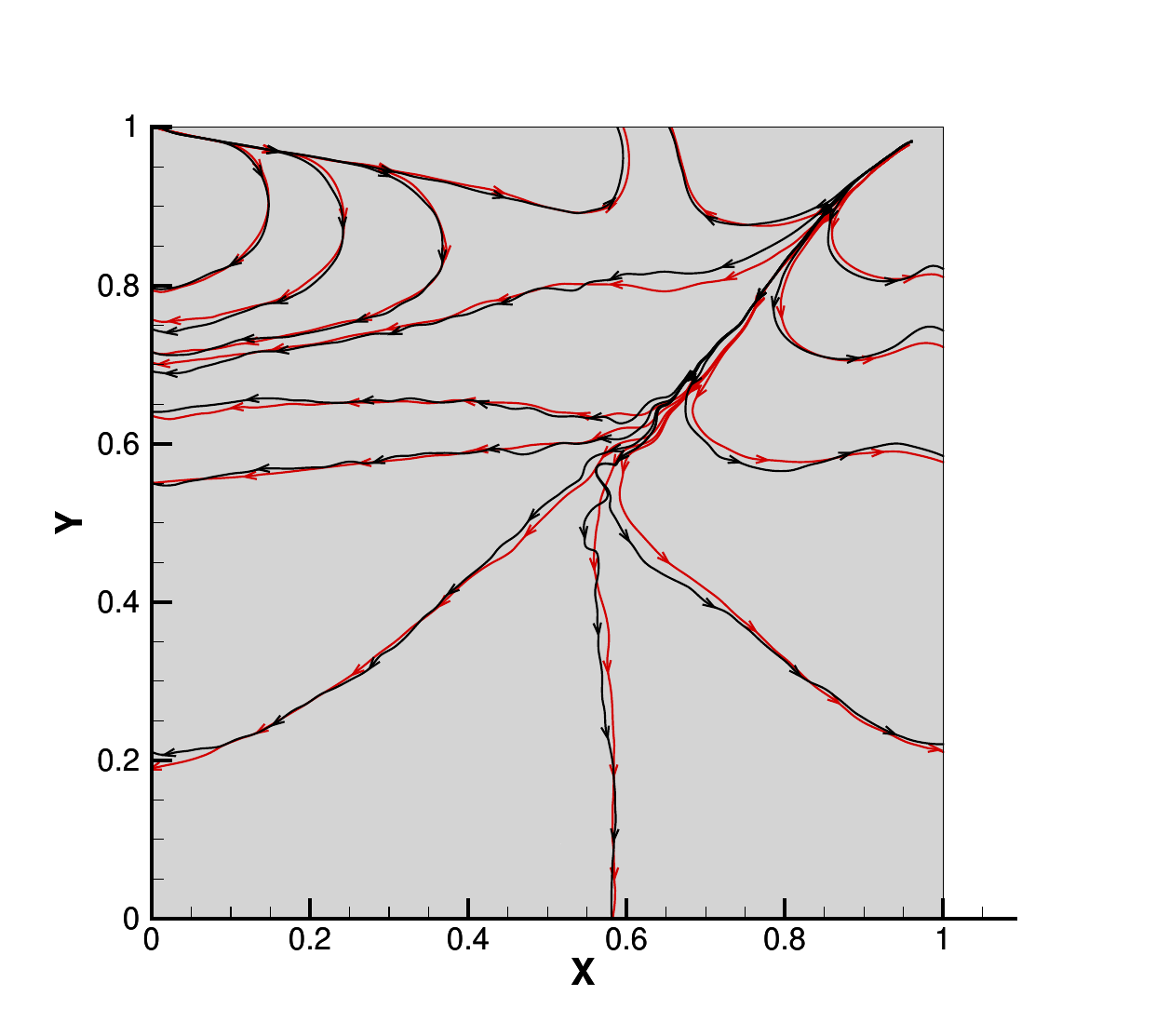}
        } \\
    \end{tabularx}
    \caption{ The contours of the translational temperature (left column), rotational temperature (middle column), and the heat flux streamlines (right column) in the lid-driven cavity flow for $\text{Kn}=0.1$ (top row) and $\text{Kn}=0.01$ (bottom row). For temperature, the background contours and black lines are obtained by DIG and DSMC methods, respectively; for the heat flux, the results obtained by DIG and DSMC methods are represented by red and black lines, respectively.}
    \label{fig:cavity01ttandtr}
\end{figure}

\begin{figure}[th]
    \centering
    \begin{tabularx}{\textwidth}{*{2}{X}} 
        \subfloat{
            \includegraphics[width=0.85\linewidth,trim={20 20 60 60},clip]{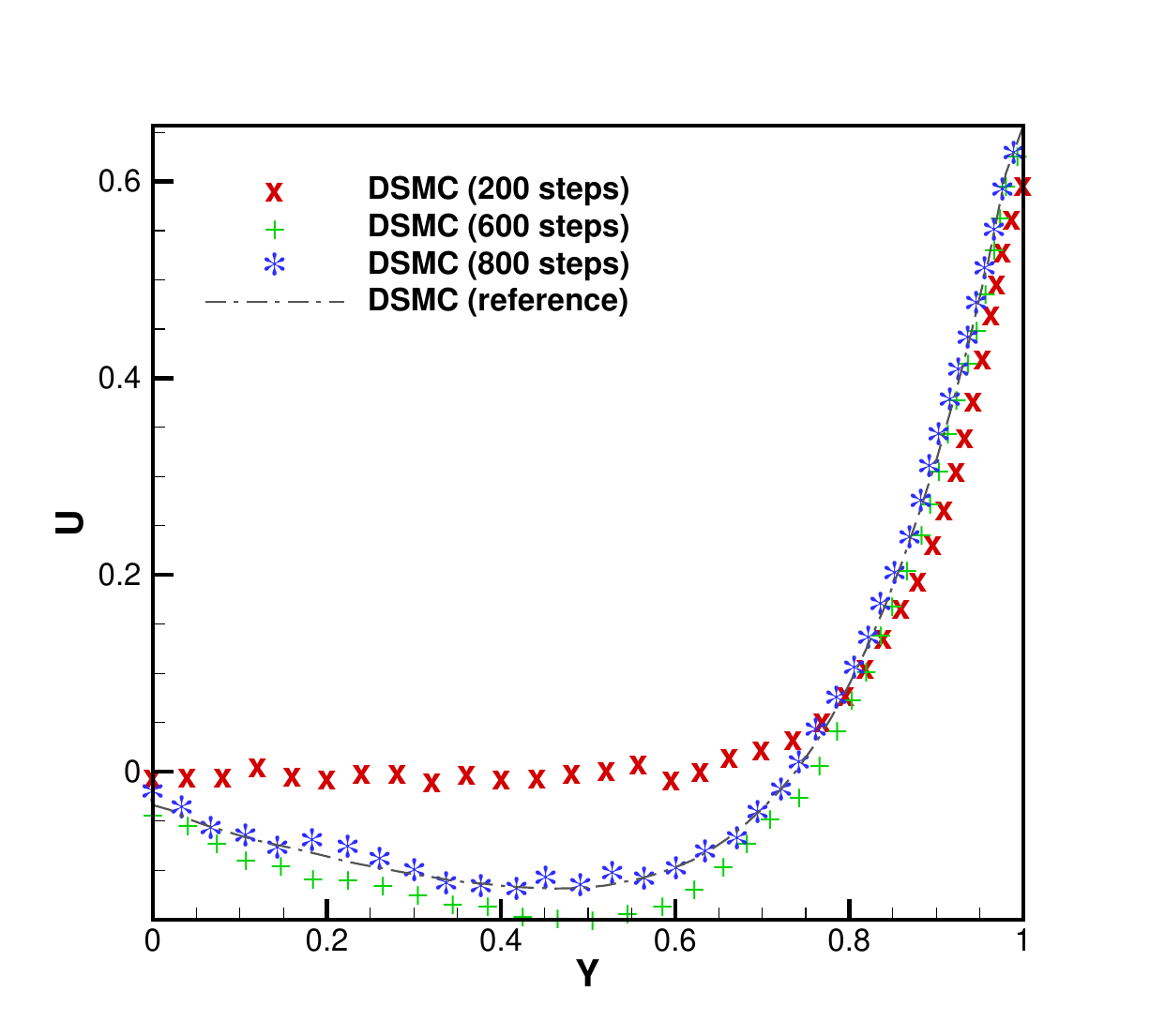}
        } &
        \subfloat{
            \includegraphics[width=0.85\linewidth,trim={20 20 60 60},clip]{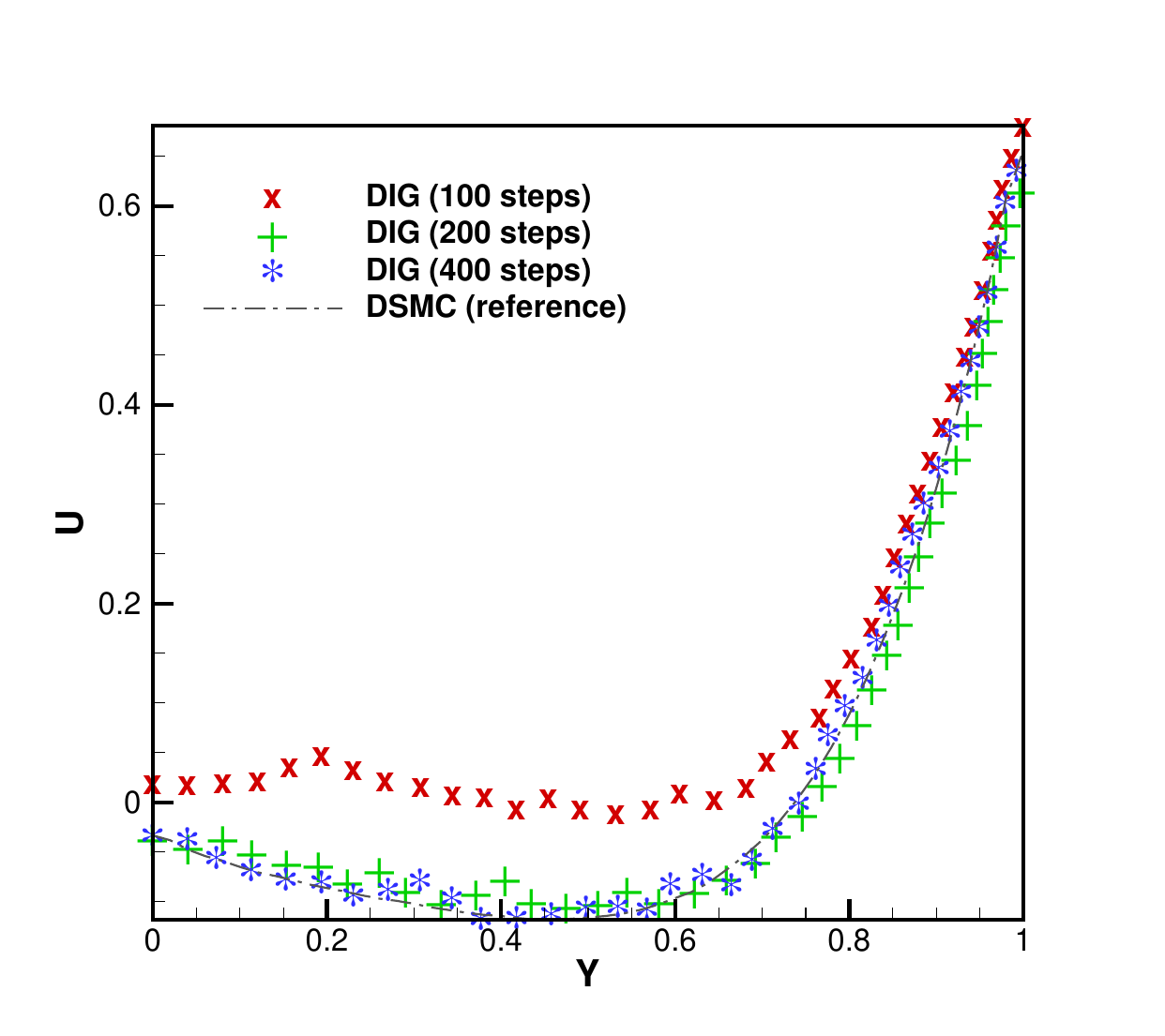}
        } \\
        \subfloat{
            \includegraphics[width=0.85\linewidth,trim={20 20 60 60},clip]{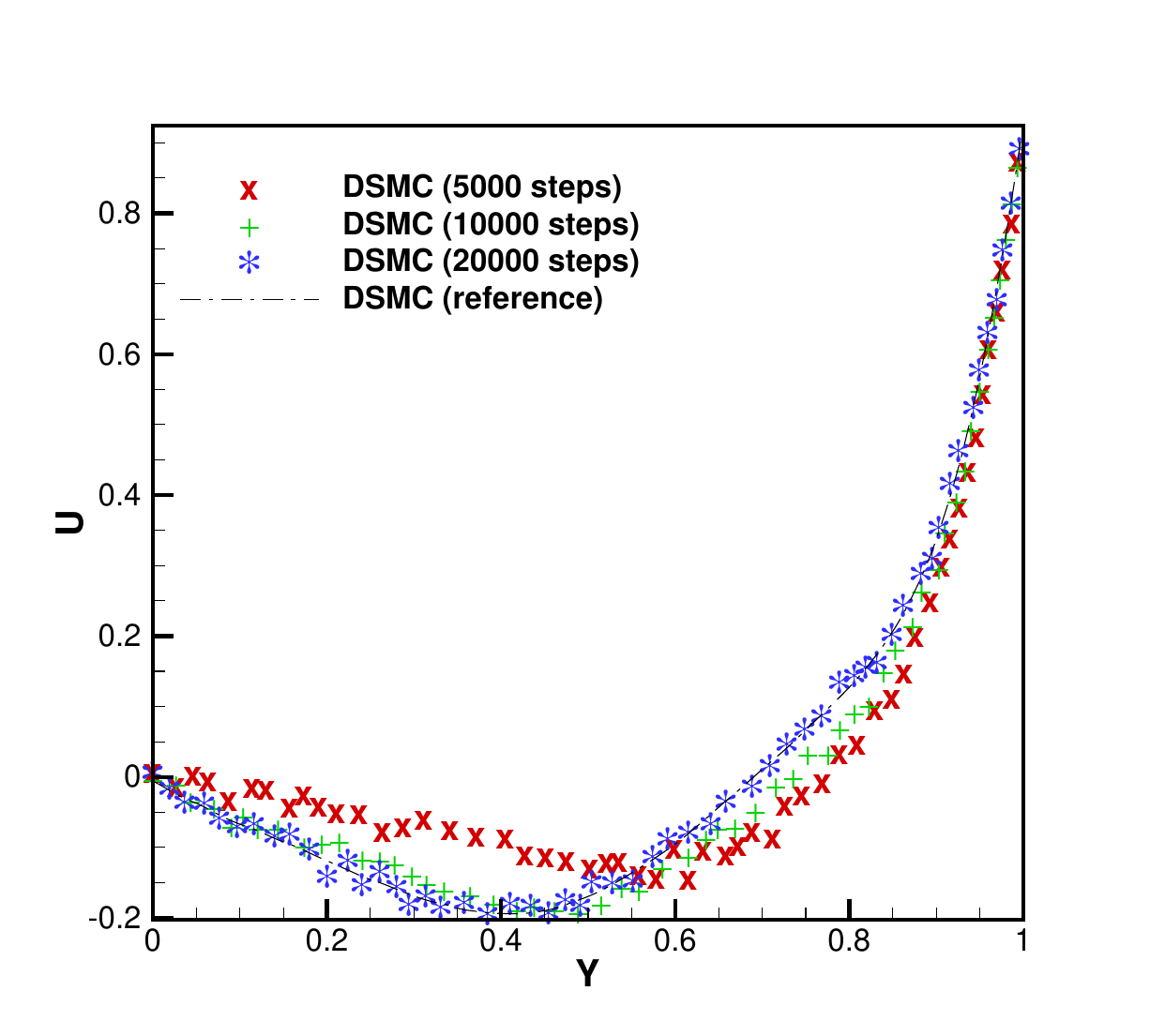}
        } &
        \subfloat{
            \includegraphics[width=0.85\linewidth,trim={20 20 60 60},clip]{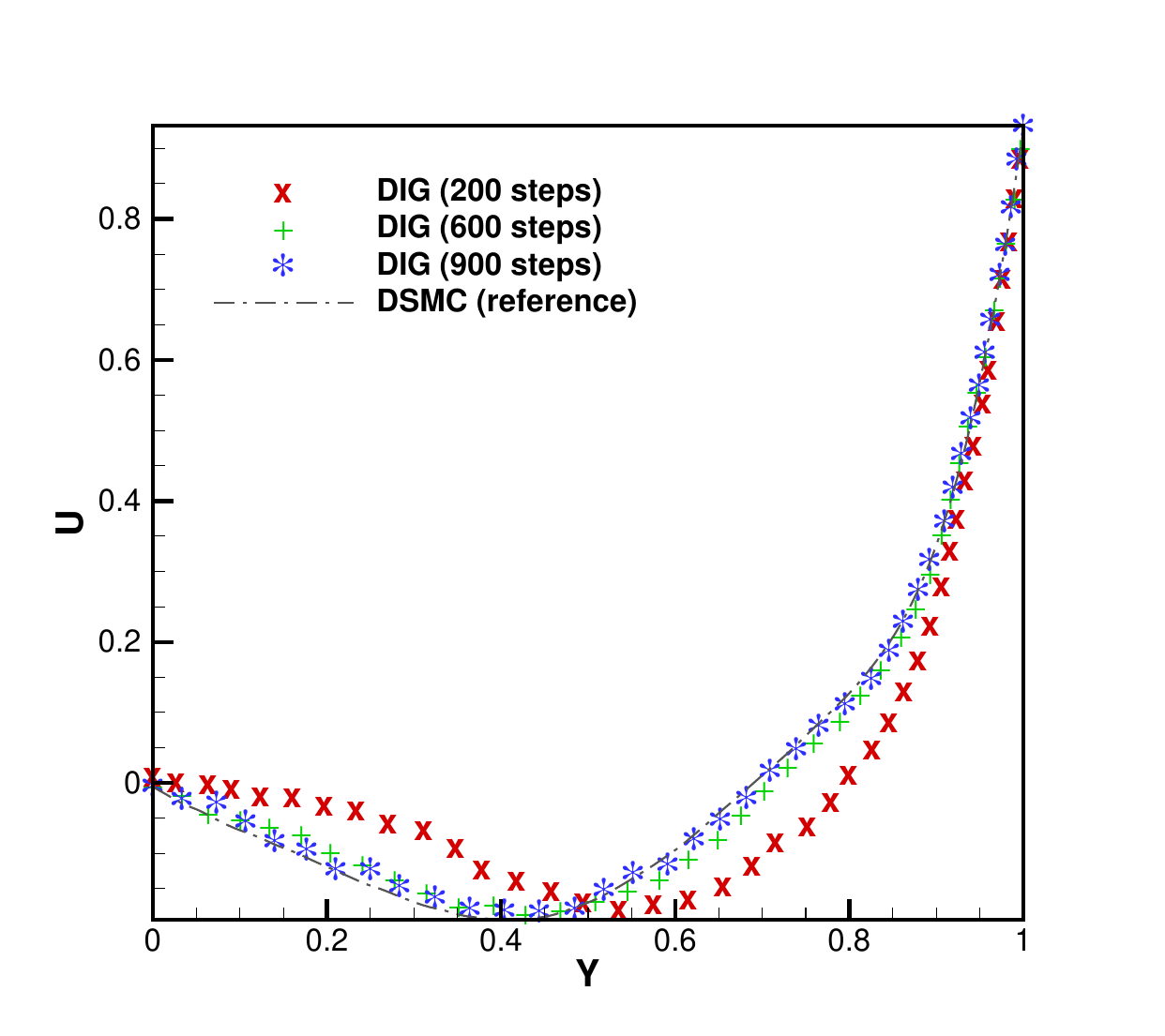}
        } \\
    \end{tabularx}
    \caption{Comparison of the horizontal velocities profiles along $x=0.5$ at different time steps between DSMC (left column) and DIG (right column) results for $\text{Kn}=0.1$ (top row) and $\text{Kn}=0.01$ (bottom row).}
    \label{fig:cavitywithefficiency}
\end{figure}

\begin{figure}[th]
    \centering
    \begin{tabularx}{\textwidth}{*{2}{X}} 
        \subfloat{
            \includegraphics[width=0.95\linewidth,trim={20 20 60 60},clip]{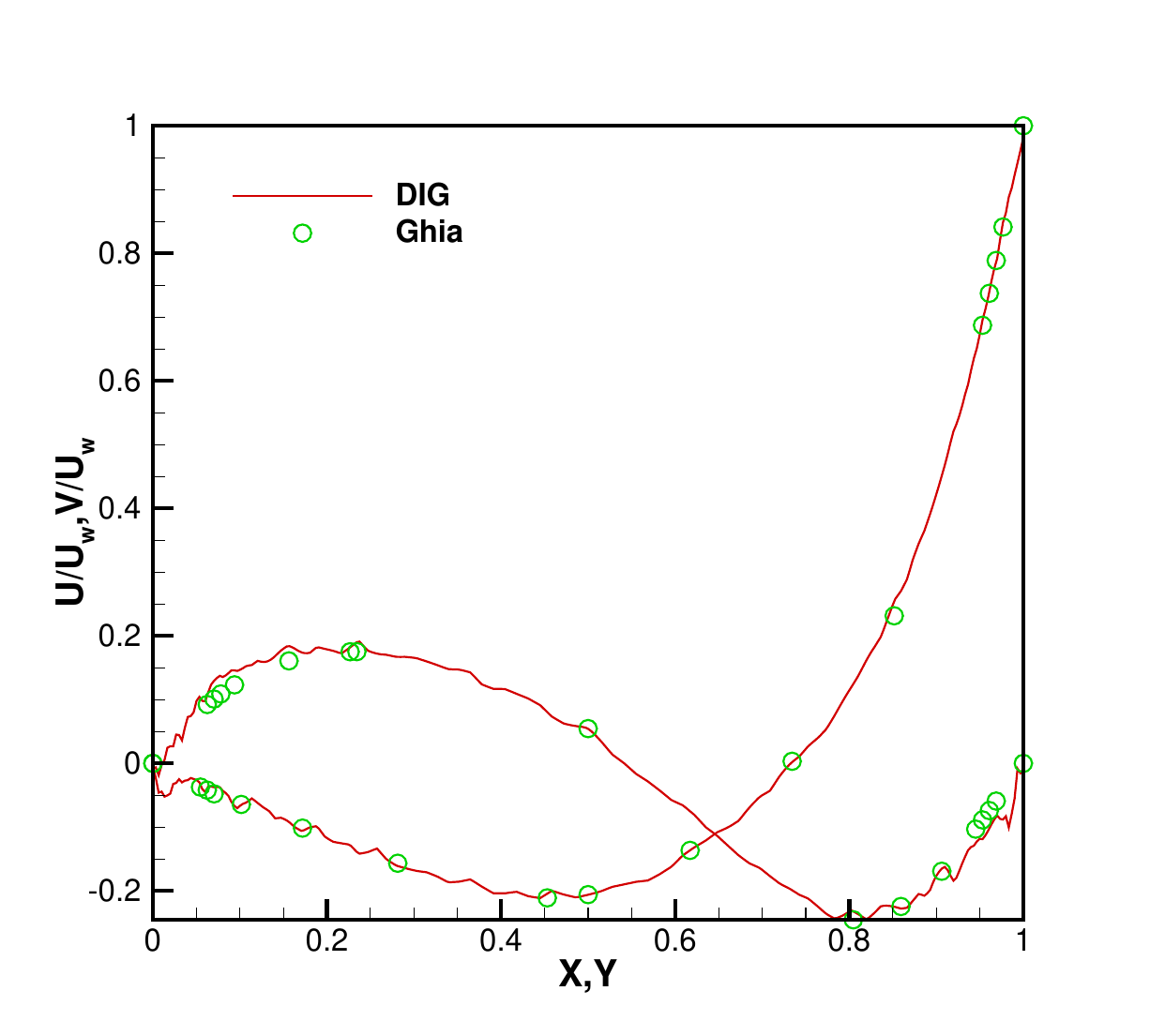}
        } &
        \subfloat{
            \includegraphics[width=0.95\linewidth,trim={20 20 60 60},clip]{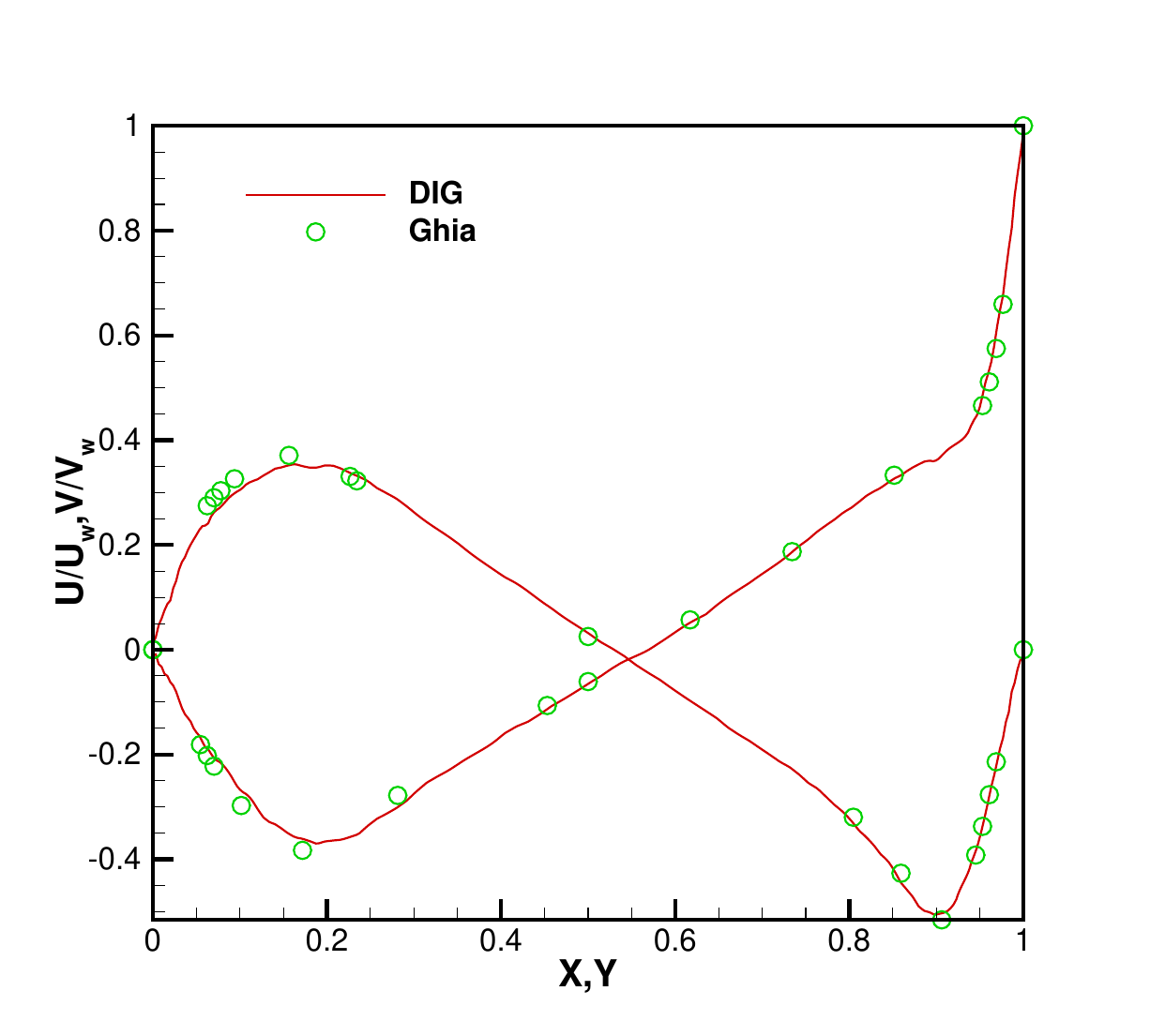}
        } \\
    \end{tabularx}
    \caption{The horizontal (vertical) velocity profiles along the centerline $x=0.5$ ($y=0.5$) of the cavity, when $\text{Kn} = 2.63 \times 10^{-3} $ (left) and $ 5.26 \times 10^{-4} $ (right), corresponding to the Reynolds number of 100 and 1000, respectively. The reference results are from Ghia~\cite{ghia-1982}.}
    \label{fig:cavitywithGhia}
\end{figure}

Figure~\ref{fig:cavity01ttandtr} presents the contours of translational and rotational temperatures, as well as the heat flux streamlines for cases with $\text{Kn} = 0.1$ and $\text{Kn} = 0.01$. When $\text{Kn} = 0.1$, both DSMC and DIG simulations utilize a uniform grid with $50 \times 50$ cells. when $\text{Kn} = 0.01$, DSMC employs a uniform grid of $300 \times 300$ cells, while DIG utilizes a non-uniform grid of $100 \times 100$ cells with refinement near the walls. The first layer of cells adjacent to the wall has a thickness of $\Delta x_{min} = 0.001$. The results obtained from the DIG simulations demonstrate excellent agreement with those obtained from DSMC.

Figure~\ref{fig:cavitywithefficiency} illustrates the evolution of the flow velocity profiles along the centerline $x=0.5$ for cases with $\text{Kn} = 0.1$ and $\text{Kn} = 0.01$. In the transition flow regime ($\text{Kn} = 0.1$), both DSMC and DIG converge to a steady-state solution within a few hundred iterations. DIG demonstrates faster convergence, requiring approximately half the number of iterations compared to DSMC. However, in the near-continuum flow regime ($\text{Kn} = 0.01$), DSMC requires a substantial number of iterations (approximately 20,000) to reach a steady-state solution. In contrast, DIG achieves steady-state convergence within only 900 iterations, highlighting the significant acceleration gained by incorporating macroscopic equation solutions to guide the particle evolution. For cases in the continuum flow regime with Knudsen numbers of $\text{Kn} = 2.63 \times 10^{-3} $ and $ 5.26 \times 10^{-4} $, the velocity profiles along the vertical and horizontal centerlines of the cavity predicted by DIG are compared to the benchmark solutions from Ghia~\cite{ghia-1982}, as shown in figure~\ref{fig:cavitywithGhia}. DIG achieves excellent agreement with the benchmark solution while maintaining a computational cost comparable to that of simulations in the near-continuum flow regime.

\begin{table}[!h]
  \centering
  \caption{Computational overhead of DIG and DSMC for lid-driven cavity flows, with computational time measured in core$\times$hours. The computational cost of DSMC simulations is unaffordable when $\text{Kn} = 2.63 \times 10^{-3}$ and $5.26 \times 10^{-4}$.}
  \begin{tabular}{cccccccccc}
    \toprule
    $\text{Kn}$ & $\text{Re}$ & $U_{w}$ & Methods & CFL & $N_{cell}$ & \multicolumn{2}{c}{Transition state} & \multicolumn{2}{c}{Steady state} \\
    \cmidrule(lr){7-8} \cmidrule(lr){9-10}
     &  &  & &  &  & steps & time & steps & time \\
    \midrule
    0.1 & 17.73 & 1.41 & DSMC & 0.2 & 50 $\times$ 50 & 800 & 0.03 & $5\times 10^{4}$ & 2.057 \\
     &  & & DIG & 0.2 & 50 $\times$ 50 & 400 & 0.04 & $3\times 10^{4}$ & 2.024 \\
    0.01 & 177.3 & 1.41 & DSMC & 0.2 & 300 $\times$ 300 & $2\times 10^{4}$ & 16.69 & $1\times 10^{5}$ & 84.28 \\
     & & & DIG & 0.5 & 100 $\times$ 100 & 900 & 0.44 & $5\times 10^{4}$ & 18.4 \\
    $2.63\times 10^{-3}$ & 100 & 0.12 & DSMC & - & - & - & - & - & - \\
     & &  & DIG & 0.5 & 150 $\times$ 150 & 1000 & 0.81 & $1\times 10^{4}$ & 7.78 \\
    $5.26\times 10^{-4}$ & 1000 & 0.25 & DSMC & - & - & - & - & - & - \\
     & & & DIG & 0.5 & 150 $\times$ 150 & 2000 & 2.65 & $1\times 10^{4}$ & 10.3 \\
    \bottomrule
    \label{tab:cavity_tab}
  \end{tabular}
\end{table}

Table~\ref{tab:cavity_tab} presents a comparison of the computational overhead of DIG and DSMC under various conditions. When $\text{Kn}=0.1$, despite the reduction in time steps required for DIG, its CPU time remains nearly equivalent to that of DSMC. This is attributed to the additional computational expense used in solving the macroscopic synthetic equations. However, as the Knudsen number decreases to $0.01$, the CPU time required by DIG in both the transition and steady states is significantly reduced, by factors of 38 and 4 respectively, compared to DSMC. Notably, for cases with $\text{Kn} = 2.63 \times 10^{-3} $ and $ 5.26 \times 10^{-4} $, DIG achieves rapid convergence to steady state. This is facilitated by the guidance provided by the synthetic equations, which also allows for computational cell sizes much larger than the gas mean free path. This highlights the fast convergence and asymptotic-preserving properties of DIG. In contrast, the computational cost of DSMC simulations in the continuum regime becomes prohibitively high.

\subsection{Hypersonic flows passing over a cylinder}

\begin{figure}[p]
    \centering
    \begin{tabularx}{\textwidth}{*{2}{X}} 
        \subfloat{
            \includegraphics[width=0.8\linewidth,trim={20 40 60 60},clip]{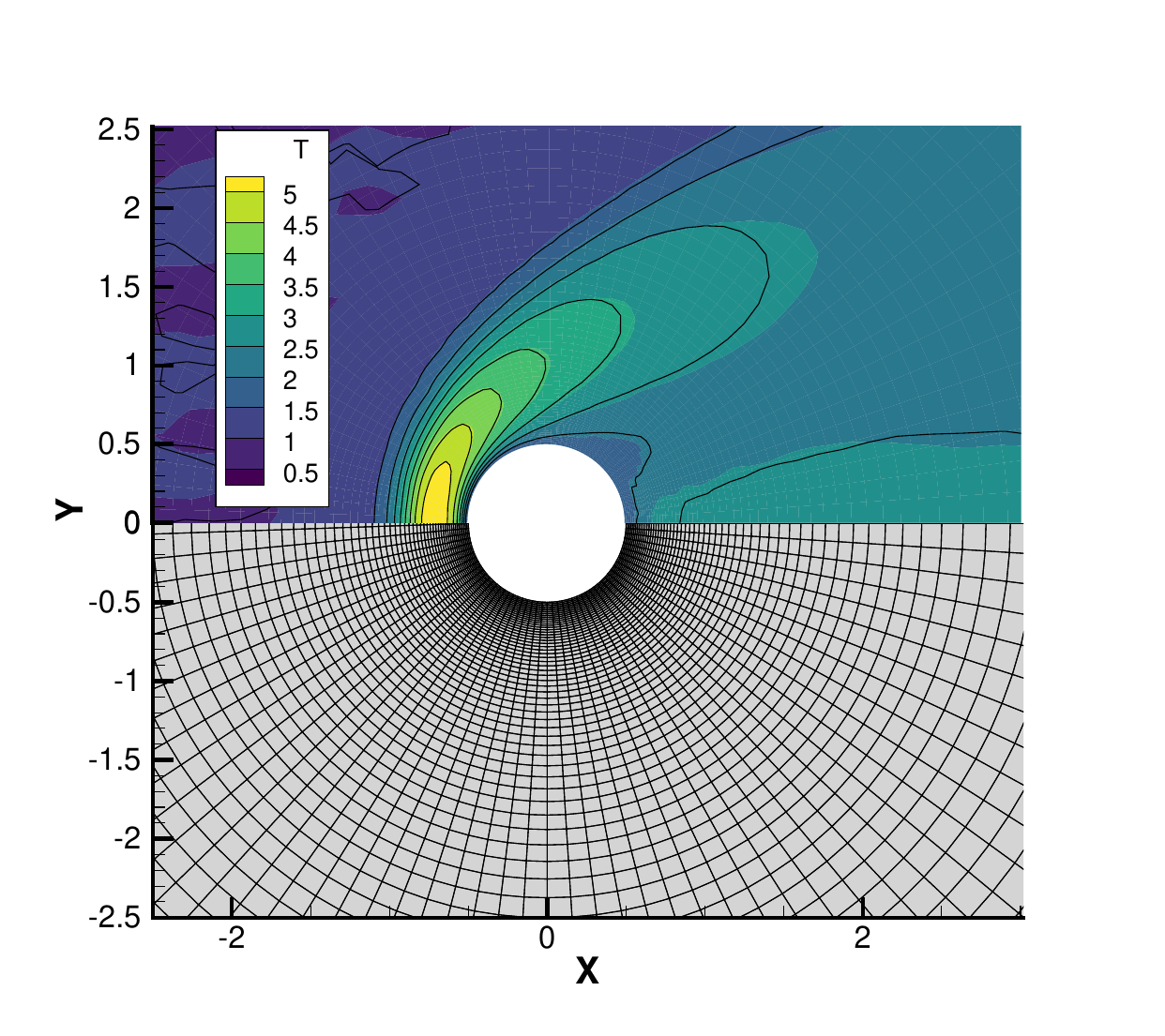}
        } &
        \subfloat{
            \includegraphics[width=0.8\linewidth,trim={20 40 60 60},clip]{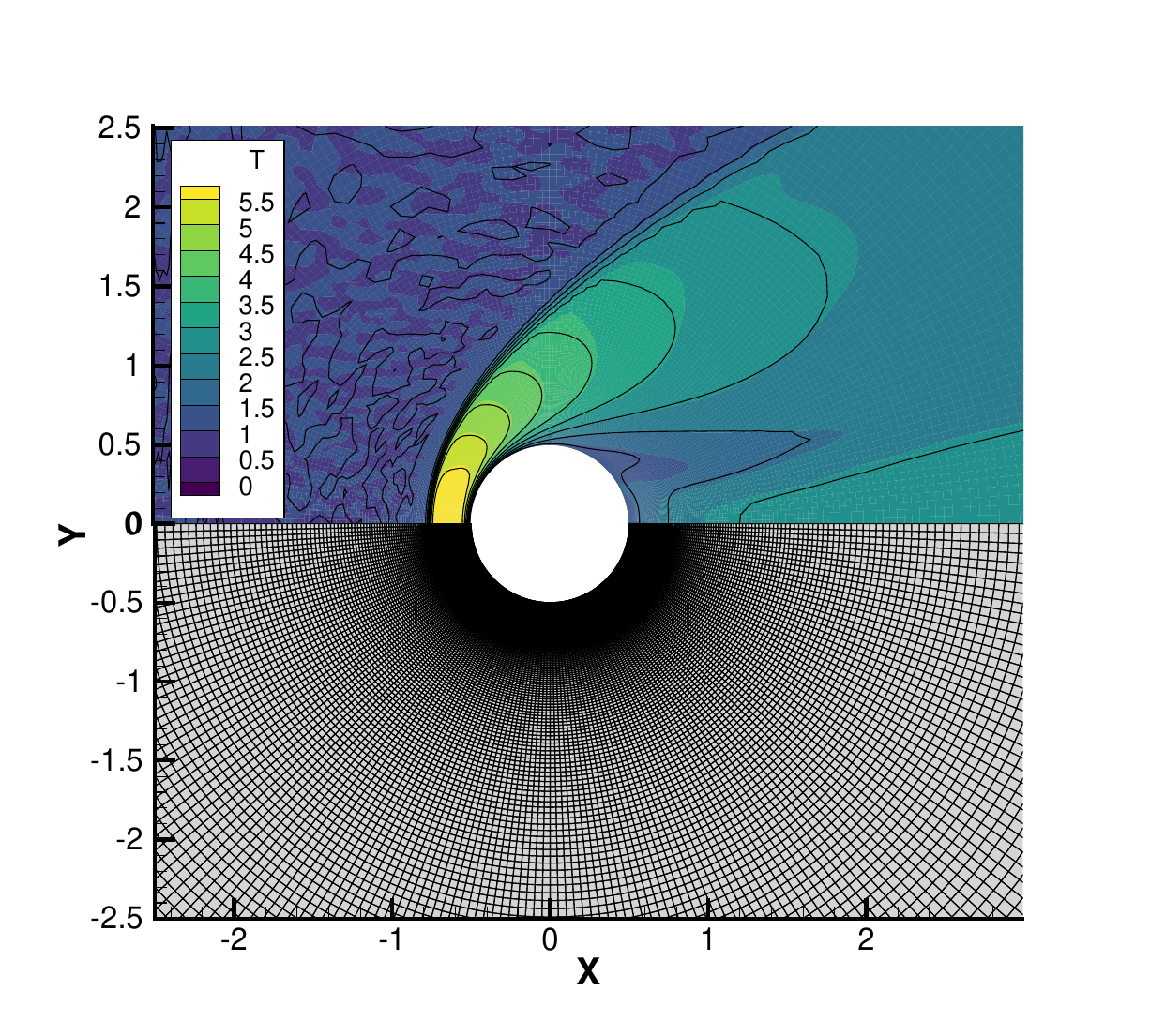}
        } \\
        \subfloat{
            \includegraphics[width=0.8\linewidth,trim={20 20 60 60},clip]{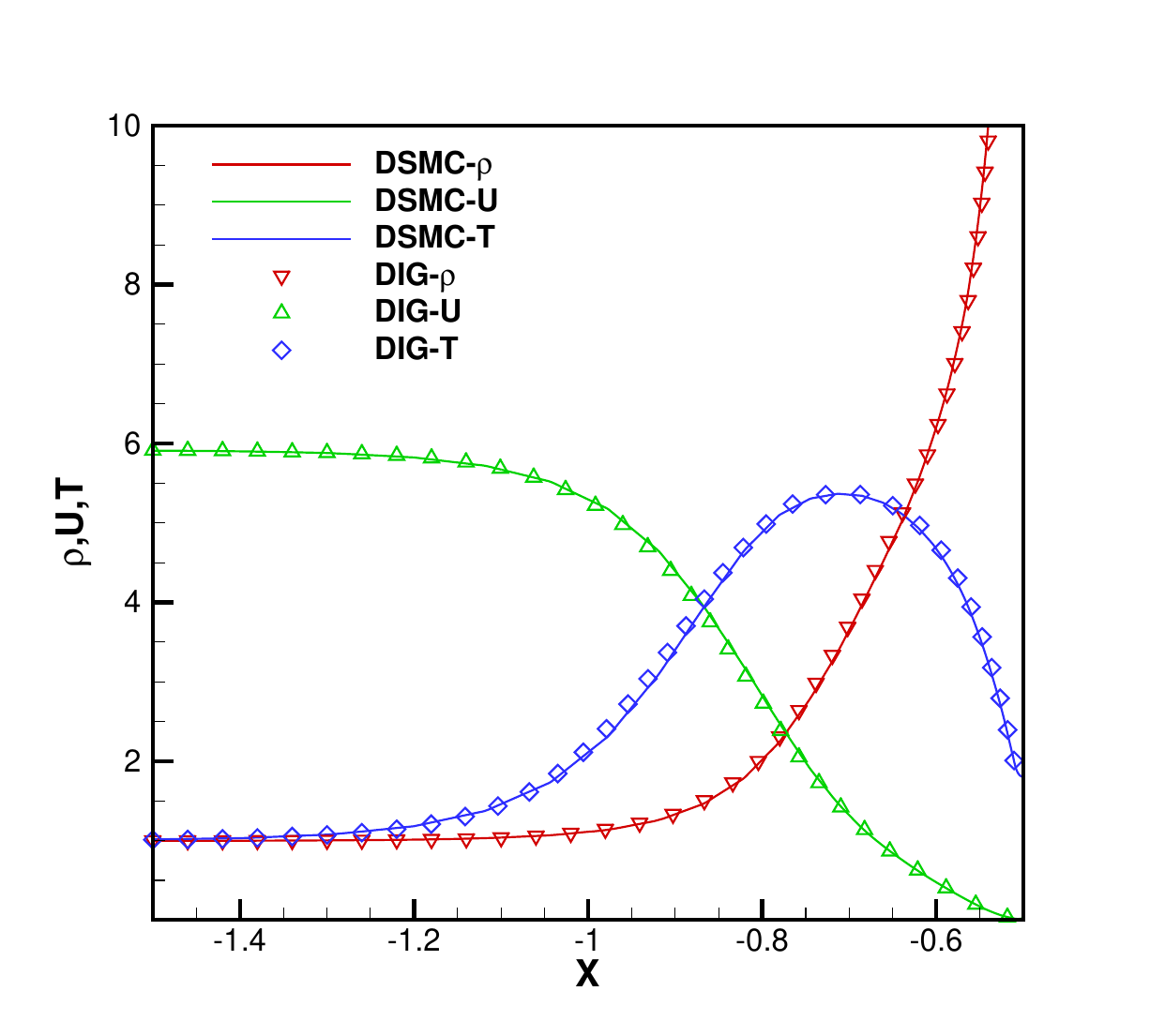}
        } &
        \subfloat{
            \includegraphics[width=0.8\linewidth,trim={20 20 60 60},clip]{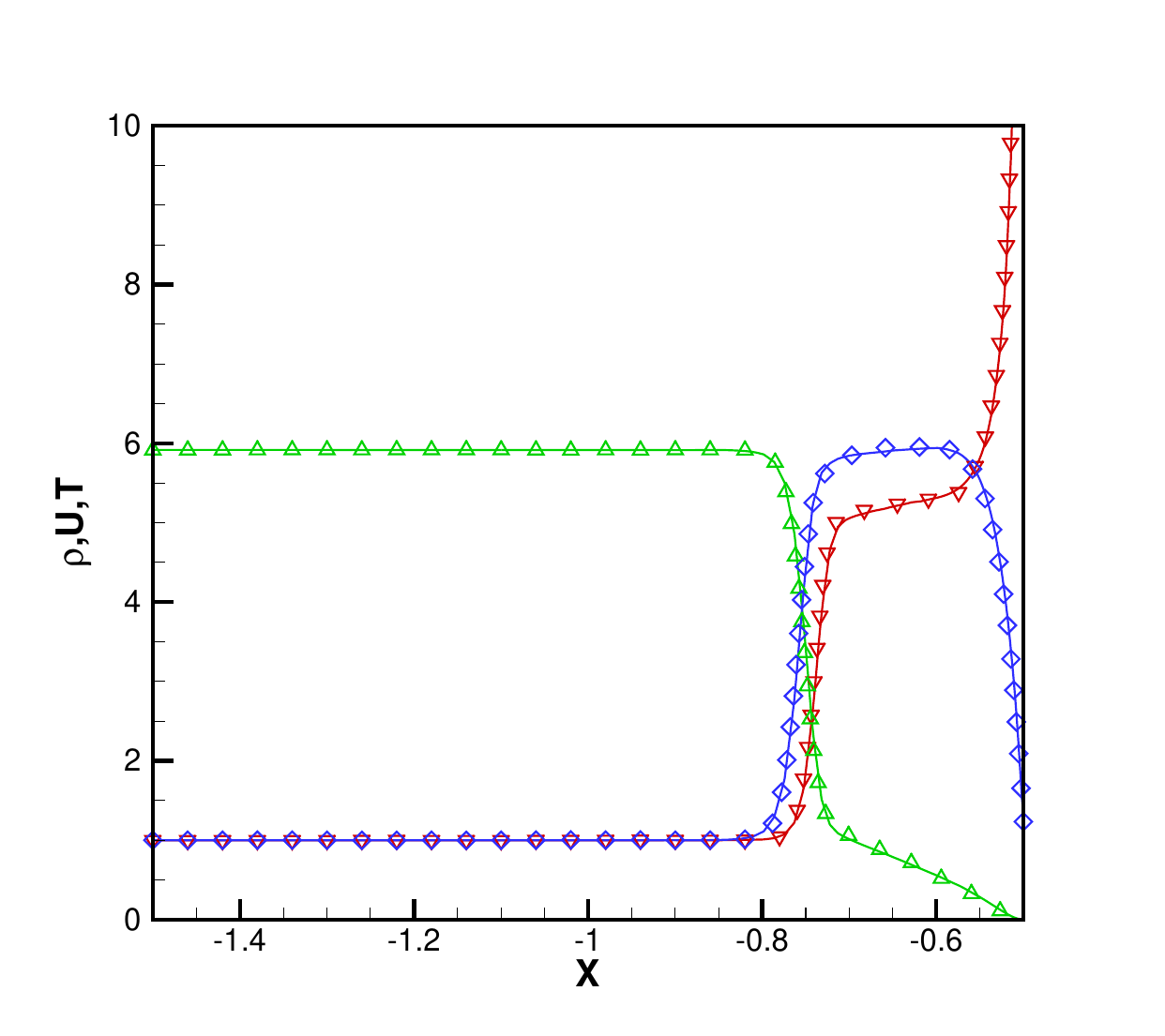}
        } \\
        \subfloat{
            \includegraphics[width=0.8\linewidth,trim={20 20 60 60},clip]{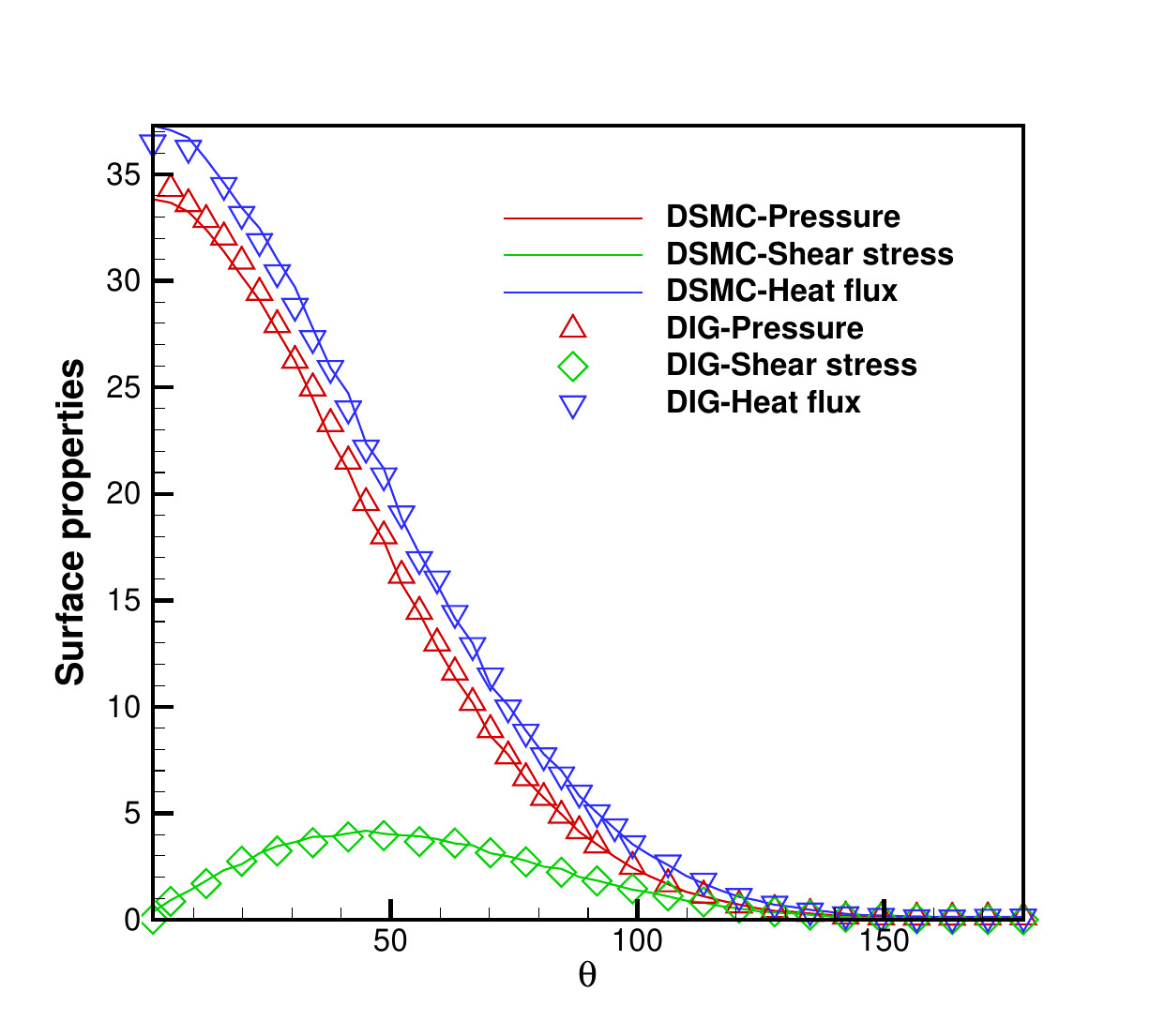}
        } &
        \subfloat{
            \includegraphics[width=0.8\linewidth,trim={20 20 60 60},clip]{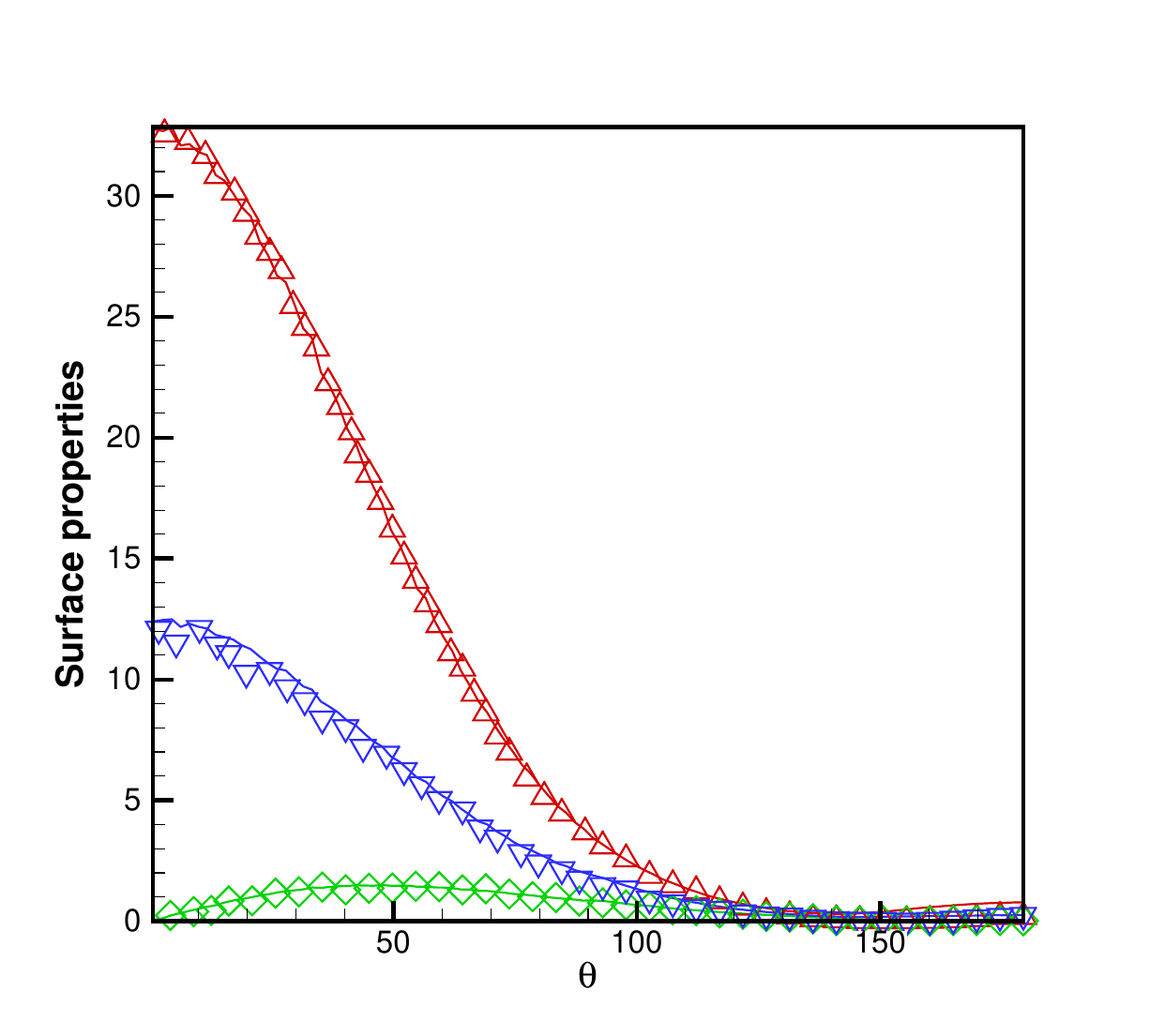}
        } \\
    \end{tabularx}
    \caption{The comparison of the steady-state results obtained by DIG and DSMC in a hypersonic flow with $\text{Ma}=5$ passing over a cylinder, when $\text{Kn} = 0.1$ (left column) and $\text{Kn} = 0.01$ (right column). (First row) The grids employed in DIG simulations and the temperature contours from both methods, where the results obtained by DIG and DSMC are presented by contours and lines respectively. (Second row) The density, velocity and temperature along the stagnation line in the windward side of the cylinder. (Third row) The pressure, shear stress and heat flux on the surface of the cylinder, where $\theta\,(^\circ)$ is the angle measured from the leading edge of the cylinder. }
    \label{fig:cylinderrhoTv_line}
\end{figure}

We conduct simulations of hypersonic flows with $\text{Ma}=5$ passing over a cylinder. The Knudsen numbers considered are $0.1$ and $0.01$, defined based on the freestream density $\rho_0$, temperature $T_0$ and the diameter of the cylinder $L_0$. The computational domain is an annular region, with an outer boundary of diameter $11L_0$ representing equilibrium free stream flow, and an inner boundary corresponding to the cylinder surface maintained at a temperature of $T_w = 1$. The entire domain is discretized into $M \times N$ structured quadrilateral meshes, with refinement near the cylinder surface. Here, $M$ and $N$ denote the number of cells in the circumferential and radial directions, respectively. As shown in Fig.~\ref{fig:cylinderrhoTv_line}, for the case with $\text{Kn} = 0.1$, both DSMC and DIG employ a grid with $M = 100$ and $N = 64$. The thickness of the first cell layer adjacent to the cylinder surface is set to $\Delta h = 0.2\lambda$. For the case with $\text{Kn} = 0.01$, the grid is refined to $M = 300,~N = 200$ in DIG and $M = 500,~N = 500$ in DSMC. An average of 200 particles are initialized in each cell, and particle velocities are sampled from an equilibrium distribution function with the same density and temperature as the freestream but zero initial velocity.

\begin{figure}[!h]
    \centering
    \includegraphics[width=0.49\linewidth,trim={20 20 60 60},clip]{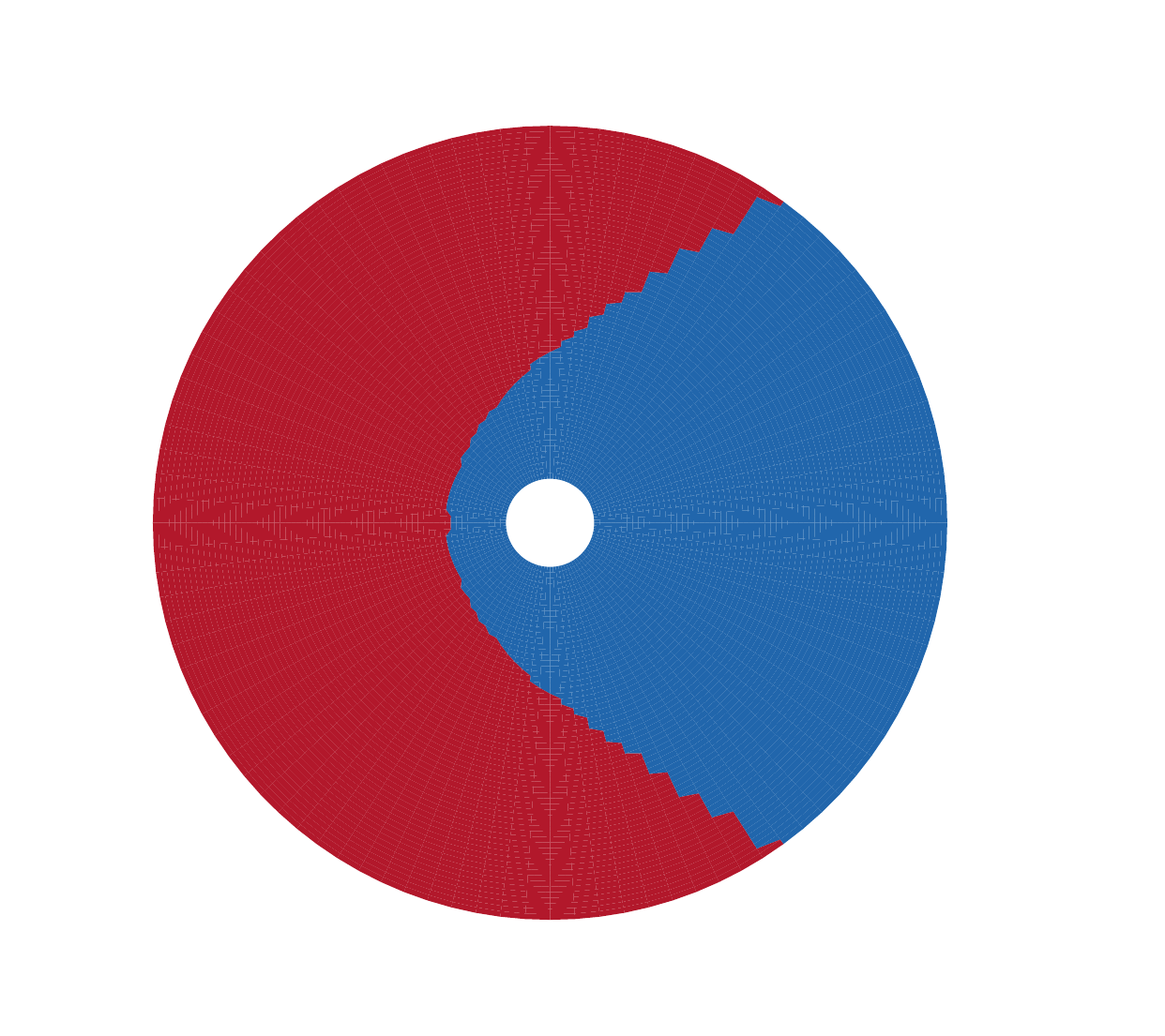}
    \includegraphics[width=0.49\linewidth,trim={20 20 60 60},clip]{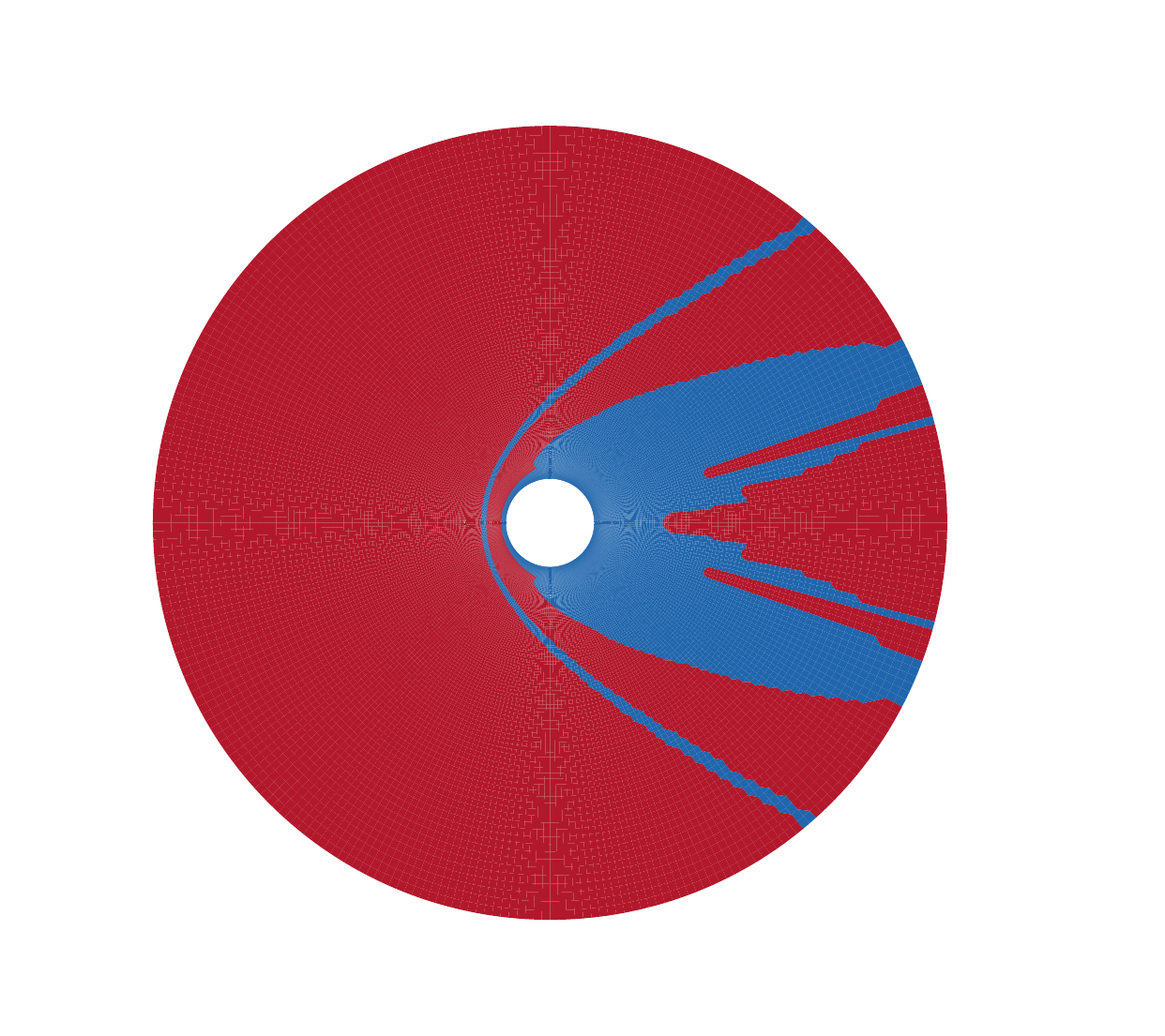}
    \caption{The adaptive regions used in DIG when $\text{Kn}=0.1$ (left) and $\text{Kn}=0.01$ (right), where higher-order terms in the red region are set to zero when solving the macroscopic synthetic equations.}
    \label{fig:knqcylinder}
\end{figure}

Figure~\ref{fig:knqcylinder} depicts the adaptive treatment utilized in the DIG method at the beginning of the transition state, using a reference Knudsen number of $\text{Kn}_{\text{ref}} = 0.01$. The red region indicates the equilibrium area, where higher-order terms are set to zero when solving the macroscopic synthesis equations. When $\text{Kn} = 0.1$, the equilibrium area accounts for approximately 44\% of the total computational domain, while increasing to around 79\% when $\text{Kn} = 0.01$.  The adaptive region is updated each time the macroscopic equations are solved, thereby dynamically differentiating between various local flow regimes. Notably, when $\text{Kn} = 0.01$, two distinct non-equilibrium regions can be clearly identified: one is the shock layer, characterized by large gradients in macroscopic properties, and the other is the downstream region, which has a longer mean free path.

\begin{figure}[!h]
    \centering
    \begin{tabularx}{\textwidth}{*{2}{X}} 
        \subfloat{
            \includegraphics[width=0.95\linewidth,trim={20 20 60 60},clip]{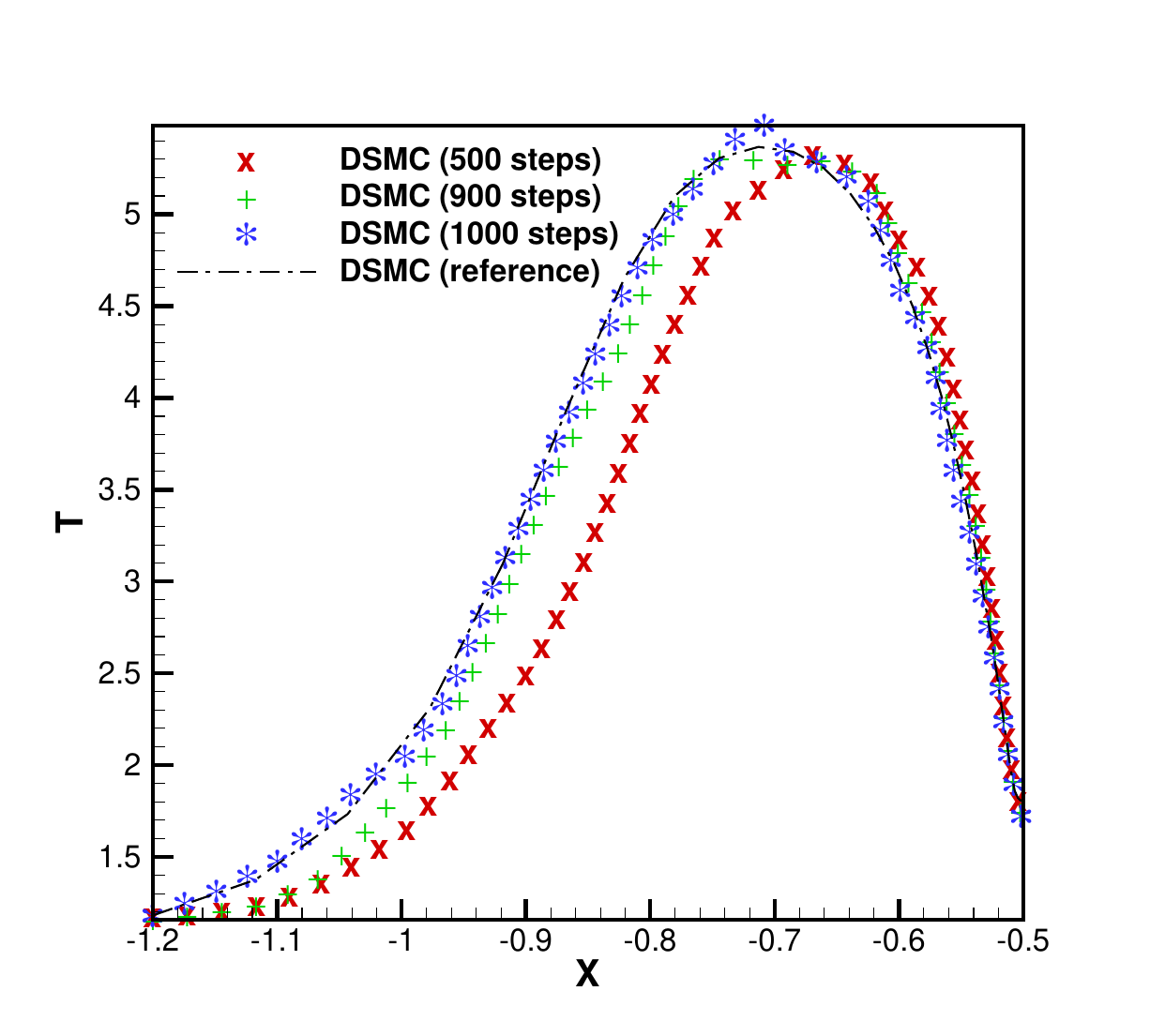}
        } &
        \subfloat{
            \includegraphics[width=0.95\linewidth,trim={20 20 60 60},clip]{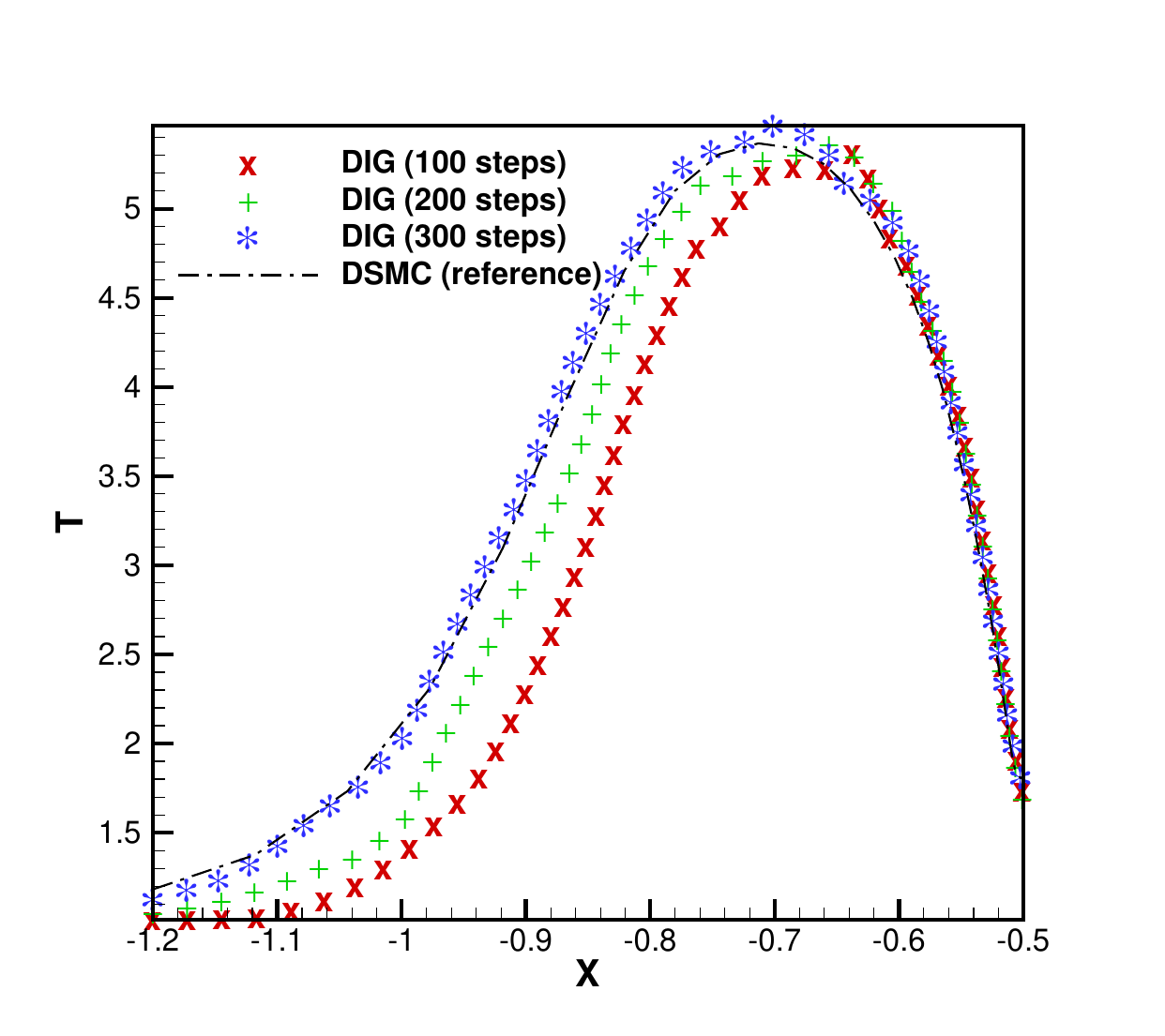}
        } \\
        \subfloat{
            \includegraphics[width=0.95\linewidth,trim={20 20 60 60},clip]{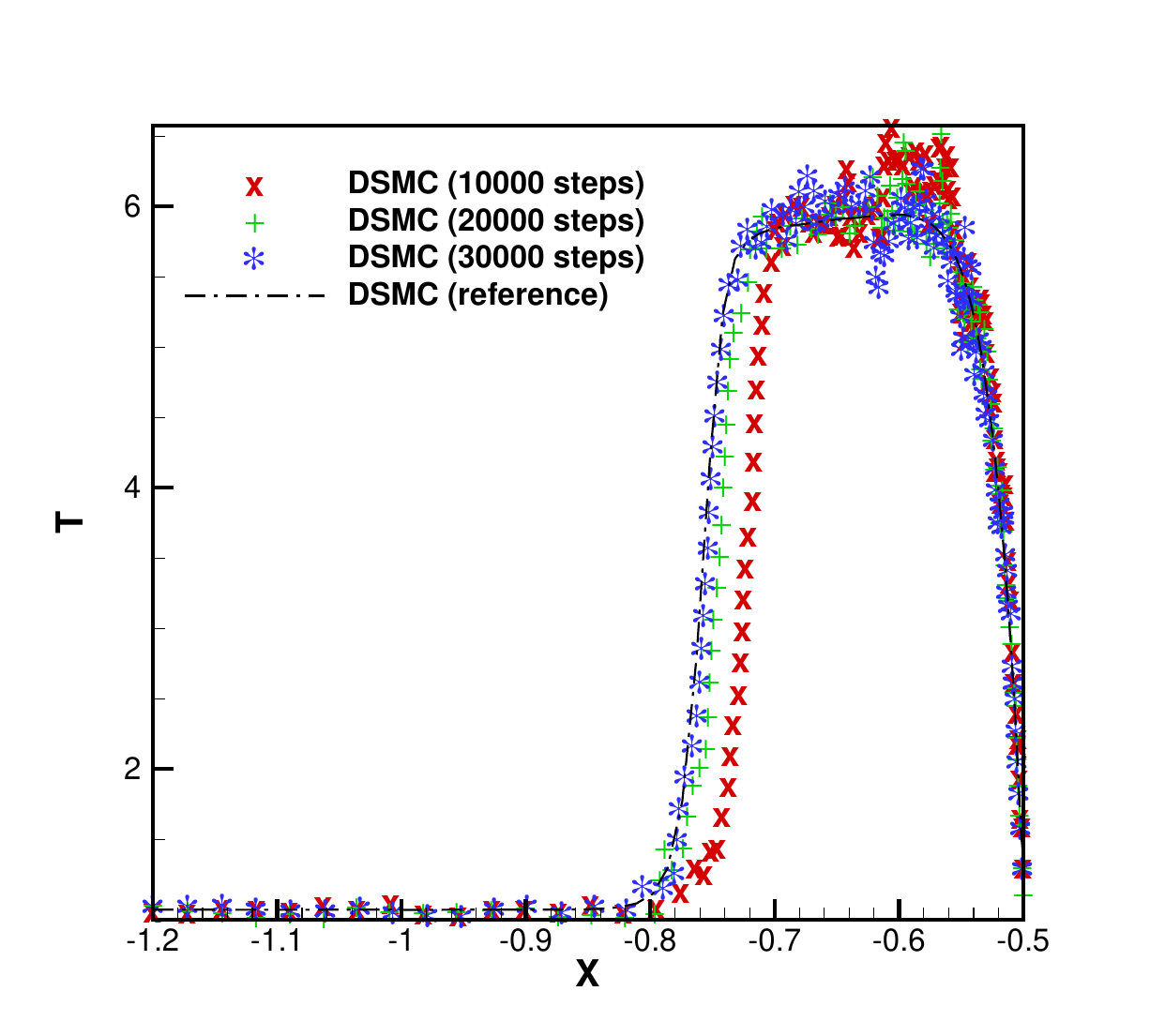}
        } &
        \subfloat{
            \includegraphics[width=0.95\linewidth,trim={20 20 60 60},clip]{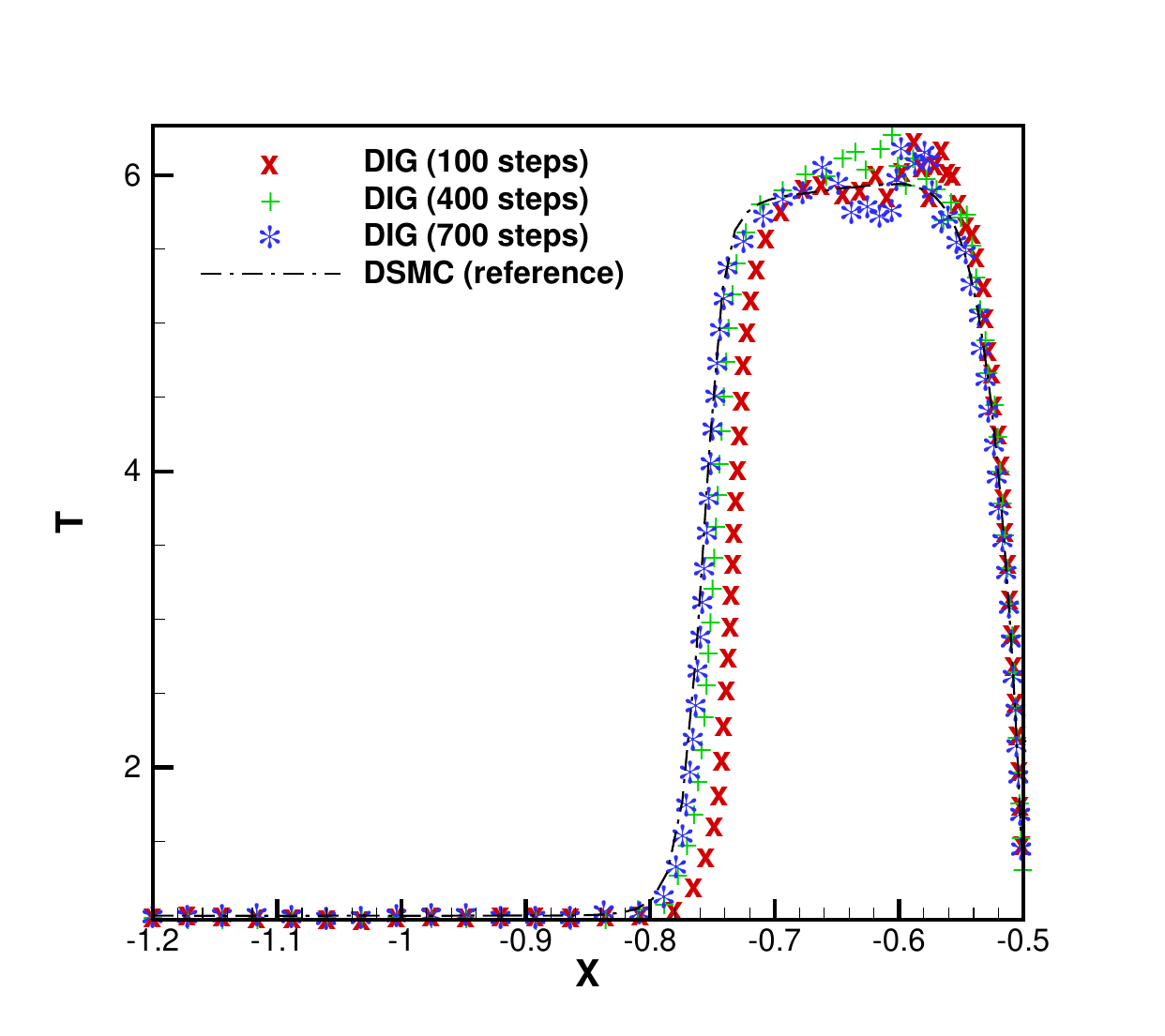}
        } \\
    \end{tabularx}
    \caption{The evolution of total temperature along the stagnation line on the windward side of the cylinder obtained by DSMC (left column) and DIG (right column), when $\text{Kn}=0.1$ (first row) and $\text{Kn}=0.01$ (second row).}
    \label{fig:cylinderteffiency}
\end{figure}

% \begin{figure}[h]
%     \centering
%     \begin{tabularx}{\textwidth}{*{2}{X}} 
%         \subfloat{
%             \includegraphics[width=0.95\linewidth,trim={20 20 60 60},clip]{figures/knqcylinder01.pdf}
%         } &
%         \subfloat{
%             \includegraphics[width=0.95\linewidth,trim={20 20 60 60},clip]{figures/knqcylinder001.pdf}
%         } \\
%     \end{tabularx}
%     \caption{The division of continuous and rarefied flow regions for $\text{Kn} = 0.1$(left) and $\text{Kn} = 0.01$(right) is shown. The green represents the continuous region, while the purple indicates the rarefied flow region.}
%     \label{fig:cylinderknq}
% \end{figure}

% Figure~\ref{fig:cylinderknq} presents the interface positions between the continuous region and the rarefied flow region under two flow conditions, $\text{Kn} = 0.1$ and $\text{Kn} = 0.01$. From these interface positions, it can be seen that as the Knudsen number decreases, the surface layer and the near-wake rarefied regions consequently diminish.

Figure~\ref{fig:cylinderrhoTv_line} compares the steady-state results obtained by DIG and DSMC, including the temperature contour, macroscopic properties along the stagnation streamline, as well as the shear stress and heat flux along the cylinder surface. Good agreements are found for all the macroscopic properties. Particularly, when $\text{Kn} = 0.01$, the cell size used in DIG is significantly coarser than the local mean free path of gas molecules, demonstrating the asymptotic-preserving property of DIG inherited from GSIS algorithm.

Figure~\ref{fig:cylinderteffiency} illustrates the evolution of total temperature along the stagnation streamline on the windward side of the cylinder, as obtained from both DIG and DSMC simulations. When $\text{Kn} = 0.1$, DSMC requires approximately 1000 iterations to reach steady state, while DIG converges within 300 iterations, demonstrating a improvement in convergence rate. As the Knudsen number decreases to 0.01, the difference in the number of steps to achieve steady state becomes more pronounced. DSMC requires 30,000 iterations, whereas DIG reaches steady state within 700 steps. 

\begin{table}[!th]
  \centering
  \caption{Computational overhead of DIG and DSMC for hypersonic flow around a cylinder, with computation time measured in core$\times$hours.}
  \begin{tabular}{ccccccccc}
    \toprule
    $\text{Kn}$  & Methods & CFL & $N_{cell}$ & \multicolumn{2}{c}{Transition state} & \multicolumn{2}{c}{Steady state} \\
    \cmidrule(lr){5-6} \cmidrule(lr){7-8}
     &  & &    & steps & time & steps & time \\
    \midrule
    0.1  & DSMC & 0.2 & 64 $\times$ 100& 1000 & 0.571 & $1\times 10^{4}$ & 5.28 \\
     &  DIG & 0.2 & 64 $\times$ 100 & 300 & 0.24 & 5000 & 3.67 \\
    0.01  & DSMC & 0.2 & 500 $\times$ 500 & $3\times 10^{4}$ & 437.2 & $8\times 10^{4}$ & 961.9 \\
     & DIG & 0.5 & 200 $\times$ 300 & 700 & 4.15 & $1\times 10^{4}$ & 57.91 \\
    \bottomrule
    \label{tab:cylinder_tab}
  \end{tabular}
\end{table}

Table~\ref{tab:cylinder_tab} summarizes the computational overhead of DIG and DSMC for all cases. When $\text{Kn}=0.1$, DIG reduces the total computational time (including both transition and steady-state phases) by approximately half compared to DSMC. However, when $\text{Kn}=0.01$, the advantage of DIG becomes even more substantial. This significant improvement is attributed to two key factors: (i) the incorporation of macroscopic equation solutions effectively guides the evolution of the particle distribution, leading to faster convergence towards the steady-state solution; (ii) the use of adaptive treatment of higher-order terms reduces the impact of statistical noise inherent in DSMC, thus allowing for a decrease in the number of time-averaged samples required to obtain smooth and accurate results. Therefore, DIG achieves a reduction in CPU time of more than two orders of magnitude compared to standard DSMC simulations in the near-continuum regime, highlighting its superior efficiency.

% \subsection{Laminar boundary layer flow passing a flat plate}

% When performing calculations in near-continuum flow regions using particle methods, significant computational time is required, and the results are often inaccurate. Therefore, It remains a challenge to achieve fast computations that recover the Navier-Stokes solution while ensuring high numerical accuracy. To further confirm the superiority of the  DIG algorithm in the multiscale numerical method, this study conducts numerical simulations of boundary layer flow passing a flat plate. The flat plate has a length of $L=100$. A gas with a density of $\rho=1.0$ and a temperature of $T=T_0$ flows passing the flat plate at a speed of $U=0.1\text{Ma}$. In this case, the Reynolds number $\text{Re}_L = U_{\infty} L/\nu = 10^5$. For the physical domain, the computational region is $[-50, 100] \times [0, 50]$ with a grid of$150 \times 50$, as shown in Figure***. The leading edge of the flat plate is positioned at $x=0$, and the minimal grid cell size is set to $\Delta x = 0.01 \text{ and } \Delta y = 0.1$. Additionally, diffuse reflection boundary conditions are applied on the plate surface, symmetric boundary conditions are used for the bottom boundary of [-50, 0], and an outflow boundary is specified on the right side. The CFL number is set to 0.5.

% The results are shown in Figure***, where the dimensionless velocities are defined as $U = U/U_{\infty}, V = V\sqrt{\text{Re}_x}/{U_{\infty}}$
\section{Conclusions and outlooks}\label{sec:conclusions}

The DIG method has been developed to accelerate DSMC simulations of polyatomic gases with internal degrees of freedom. This algorithm exhibits fast convergence and asymptotic-preserving properties in multiscale gas flow problems, while maintaining the fundamental framework of the standard DSMC method but incorporating a macroscopic solver intermittently. At specific intervals during the DSMC time-stepping, the macroscopic synthetic equations are solved, and the particle distribution within the DSMC framework is adjusted based on the solutions of these macroscopic equations. Furthermore, an adaptive treatment is implemented for the constitutive relations used in the macroscopic solver. Higher-order terms extracted from DSMC are applied exclusively in rarefied regions where the local Knudsen number exceeds a predefined reference value. This adaptive approach enhances stability and reduces the required sampling size for obtaining accurate and smooth results. It is important to note that while DIG allows for significantly larger spatial cell sizes compared to gas mean free path, the time step must still remain smaller than the mean collision time, as in standard DSMC simulations.

Through numerical simulations of nitrogen gas flows, including lid-driven cavity flows and hypersonic flows past a cylinder, across a wide range of Knudsen numbers, the polyatomic DIG method has demonstrated its capability for rapid and accurate solution in all flow regimes. In particular, it achieves a reduction in computational time of over two orders of magnitude compared to standard DSMC in the near-continuum flow regime, and maintains computational efficiency even in the continuum flow regime where DSMC becomes prohibitively expensive.

Building upon its minimally modified framework of standard DSMC, DIG can be feasibly extended to simulate complex gas flows involving intricate processes such as gas mixtures, chemical reactions, and even plasma phenomena. Moreover, the DIG algorithm can be effectively adapted to study the interactions between rarefied and turbulent gas flows~\cite{tian-2024}. This adaptability, coupled with its significant computational advantages, positions DIG as a promising approach for addressing challenging multiscale gas flow problems.

\section*{Acknowledgments}

This work is supported by the National Natural Science Foundation of China (no. 12202177), the Guangdong Basic and Applied Basic Research Foundation (no. 2024A1515011393), the University Stable Support Research Funding of Shenzhen (no. 20231116171911001).

% \appendix

\bibliographystyle{elsarticle-num}
\bibliography{ref}

\end{document}